\documentclass[10pt,journal,compsoc]{IEEEtran}
\ifCLASSOPTIONcompsoc
  \usepackage[nocompress]{cite}
\else
  \usepackage{cite}
\fi
\ifCLASSINFOpdf
\else
\fi
\usepackage{amsmath,amsfonts}
\usepackage{algorithmic}
\usepackage{algorithm}
\usepackage{array}
\usepackage{textcomp}
\usepackage{stfloats}
\usepackage[hyphens]{url}
\usepackage[hidelinks]{hyperref}
\usepackage{multicol}
\usepackage{multirow}
\usepackage{rotating}
\usepackage{arydshln}
\usepackage{verbatim}
\usepackage{graphicx}
\usepackage{cite}
\usepackage{booktabs}
\usepackage{threeparttable}
\usepackage{makecell}
\usepackage[edges]{forest}
\usepackage{tikz}
\usepackage{enumitem}

\usetikzlibrary{positioning}
\usetikzlibrary{shapes.geometric}
\usetikzlibrary{arrows}

\usepackage{hyperref}
\usepackage[capitalize,noabbrev]{cleveref}
\tikzset{
    basic/.style  = {draw, text width=2cm, drop shadow, font=\sffamily, rectangle},
    bigbox/.style={draw=black, rounded corners=5pt, fill=gray!20, text width=10.5cm, minimum height=0.8cm, align=center},
    midbox/.style={draw=black, rounded corners=5pt, fill=white, text width=4.5cm, minimum height=1cm, align=center},
    smallbox/.style={draw=black, rounded corners=5pt, fill=white, text width=3.5cm, minimum height=0.6cm, align=center}
    edge from parent/.style={draw=black, edge from parent fork right}
    }

\begin{document}

\title{Few-shot Molecular Property Prediction: A Survey}

\author{Zeyu~Wang,~Tianyi~Jiang,~Huanchang~Ma,~Yao~Lu,~\IEEEmembership{Student Member,~IEEE},~Xiaoze~Bao,~Shanqing~Yu,\\~Qi~Xuan,~\IEEEmembership{Senior Member,~IEEE}, ~Shirui~Pan,~\IEEEmembership{Senior Member,~IEEE},~Xin~Zheng
\IEEEcompsocitemizethanks{
\IEEEcompsocthanksitem Z. Wang is with the Institute of Cyberspace Security, College of Information Engineering, Zhejiang University of Technology, 310023, Hangzhou, China; Binjiang Institute of Artificial Intelligence, Zhejiang University of Technology, 310056, Hangzhou, China; the School of Information and Communication Technology, Griffith University, Southport, QLD 4215, Australia (E-mail: Vencent\_Wang@outlook.com).\protect
\IEEEcompsocthanksitem T. Jiang, S. Yu, and Q. Xuan are with the Institute of Cyberspace Security, College of Information Engineering, Zhejiang University of Technology, 310023, Hangzhou, China; Binjiang Institute of Artificial Intelligence, Zhejiang University of Technology, 310056, Hangzhou, China (E-mail: josieyi0319@163.com; yushanqing@zjut.edu.cn; xuanqi@zjut.edu.cn).\protect
\IEEEcompsocthanksitem H. Ma is with the School of Information Science and Technology, Northeast Normal University, Changchun, Jilin, 130117, China (E-mail: mahc569@nenu.edu.cn).\protect
\IEEEcompsocthanksitem Y. Lu is with the Institute of Cyberspace Security, College of Information Engineering, Zhejiang University of Technology, Hangzhou 310023, China, with the Binjiang Institute of Artificial Intelligence, Zhejiang University of Technology, Hangzhou 310056, China, also with the Centre for Frontier AI Research, Agency for Science, Technology and Research, Singapore 138632 (e-mail: yaolu.zjut@gmail.com).\protect
\IEEEcompsocthanksitem X. Bao is with the College of Pharmaceutical Science \& Collaborative Innovation Center of Yangtze River Delta Region Green Pharmaceuticals, Zhejiang University of Technology, 310014, Hangzhou, China; Binjiang Institute of Artificial Intelligence, Zhejiang University of Technology, 310056, Hangzhou, China (E-mail: baoxiaoze@zjut.edu.cn).\protect
\IEEEcompsocthanksitem X. Zheng, S. Pan are with the School of Information and Communication Technology, Griffith University, Southport, QLD 4215, Australia (e-mail: xin.zheng@griffith.edu.au; s.pan@griffith.edu.au).\protect
\IEEEcompsocthanksitem Corresponding author: Qi Xuan and Xin Zheng.
}
}



\markboth{Journal of \LaTeX\ Class Files,~Vol.~14, No.~8, August~2021}%
{Shell \MakeLowercase{\textit{et al.}}: A Sample Article Using IEEEtran.cls for IEEE Journals}

\IEEEpubid{0000--0000/00\$00.00~\copyright~2021 IEEE}

\IEEEtitleabstractindextext{%
\begin{abstract}
AI-assisted molecular property prediction has become a promising technique in early-stage drug discovery and materials design in recent years. However, due to high-cost and complex wet-lab experiments, real-world molecules usually experience the issue of scarce annotations, leading to limited labeled data for effective supervised AI model learning.
In light of this, few-shot molecular property prediction (FSMPP) has emerged as an expressive paradigm that enables learning from only a few labeled examples. Despite rapidly growing attention, existing FSMPP studies remain fragmented, without a coherent framework to capture methodological advances and domain-specific challenges. In this work, we present the first comprehensive and systematic survey of few-shot molecular property prediction. We begin by analyzing the few-shot phenomenon in molecular datasets and highlighting two core challenges: (1) \textit{cross-property generalization under distribution shifts}, where each task corresponding to each property, may follow a different data distribution or even be inherently weakly related to others from a biochemical perspective, requiring the model to transfer knowledge across heterogeneous prediction tasks, and (2) \textit{cross-molecule generalization under structural heterogeneity}, where molecules involved in different or same properties may exhibit significant structural diversity, making model difficult to achieve generalization. 
Then, we introduce a unified taxonomy that organizes existing methods into data, model, and learning paradigm levels, reflecting their strategies for extracting knowledge from scarce supervision in few-shot molecular property prediction.
Next, we compare representative methods, summarize benchmark datasets and evaluation protocols.
In the end, we identify key trends and future directions for advancing the continued research on FSMPP. The repository of papers, code, and datasets is at~\url{https://github.com/Vencent-Won/Awesome-Literature-on-Few-shot-Molecular-Property-Prediction}.
\end{abstract}

}
\maketitle

\section{Introduction}
\begin{figure}[!t]
    \centering
    \includegraphics[width=\linewidth]{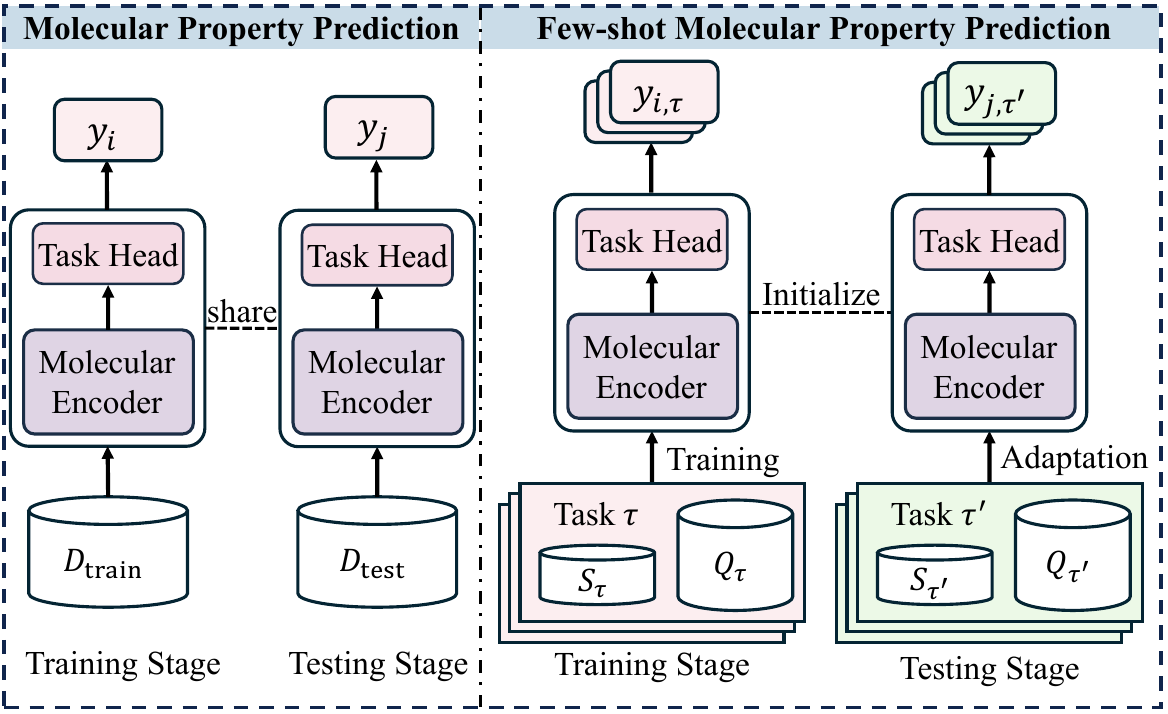}
    \caption{Difference between MPP and FSMPP, where $y_i$ is the prediction of molecule $x_i$, $y_{i,\tau}$ is the prediction under task $\tau$, $S_\tau$, $Q_\tau$ denotes the support set and query set of task $\tau$. With $|D_{\text{train}}| \gg |S_\tau|$, FSMPP requires generalization across both molecules and tasks under very limited supervision.}
    \label{fig: scenaories comparison}
\end{figure}
\IEEEPARstart{M}olecular property prediction (MPP) aims to accurately estimate the physicochemical properties and biological activities of molecules, which plays a critical role in early stages of drug discovery~\cite{pan2022deep}, such as hit identification~\cite{bleicher2003hit,turon2023first}, lead optimization~\cite{lewis2005general, dara2022machine}, and drug repurposing~\cite{vamathevan2019applications}. Traditionally, MPP was conducted with wet-lab experiments that not only require large amounts of reagents and expensive instruments but also are time-consuming. In recent years, with the rapid progress of artificial intelligence (AI), data-driven methods~\cite{walters2020applications,li2021computational} have emerged as an efficient alternative for MPP. These approaches learn molecular representations by exploiting intrinsic structural information and follow the paradigm of optimizing models on labeled training data to generalize to unseen molecules, thereby accelerating drug discovery and reducing resource consumption.

Early studies in AI-assisted MPP often relied on feature engineering, such as molecular descriptors~\cite{hu2007targeting,nocedo2019modeling} and molecular fingerprints~\cite{zoffmann2019machine,sheng2023accelerating}. These predefined features can be combined with traditional machine learning algorithms for classification or regression tasks. To break the information capture limitations of such conventional chemical heuristic methods, deep learning-based technologies have achieved promising progress in MPP tasks~\cite{atz2021geometric,li2022deep}, where molecules are always represented as Simplified Molecular Input Line Entry System (SMILES) strings~\cite{weininger1988smiles}, molecular graphs~\cite{rouvray1991chemical}, or 3D conformations~\cite{fang2022geometry}. Subsequently, sequence models~\cite{wang2019smiles}, graph neural networks~\cite{wieder2020compact,heid2023chemprop}, and multi-modal~\cite{wang2024multi} learning methods are implemented to extract the underlying features of molecules with supervision signals from labeled molecular properties, achieving state-of-the-art molecular property prediction performance.

\begin{figure}[!t]
    \centering
    \includegraphics[width=\linewidth]{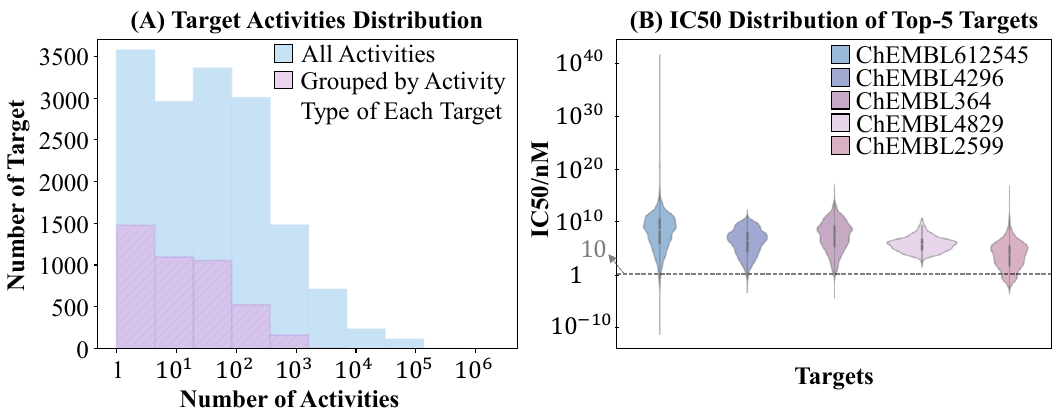}
    \caption{Data statistics of ChEMBL. (A) Distribution of activity annotations per target. The x-axis shows the number of activity annotations (log scale), and the y-axis indicates the number of targets, where a target refers to a biomolecular entity (such as a protein) that can be regarded as a specific prediction task. Two versions are shown: all activities (blue), removing outliers/duplicates and grouping by the main activity type of each target (purple). (B) IC50 distribution of the top-5 targets with the most annotations. The violin plots illustrate the spread and density of IC50 values, where IC50 is a widely used pharmacological indicator of compound activity (lower values indicate stronger activity) for each target. IC50 10 is the binary threshold.}
    \label{fig: Data Statistics}
\end{figure}

\noindent \textbf{Core challenges: scarce \& low-quality molecular annotations.}
However, due to the high-cost and complex wet-lab experiments, real-world molecules usually experience the issue of scarce annotations, leading to limited labeled molecular properties for effectively supervising deep learning models. Although certain molecular property annotations have been reported in the literature, accurate molecular annotations with high quality are still challenging~\cite{wu2024instructor, wang2025molecular,ryu2019bayesian,ijcai2024p0621}. 
For a better understanding of the molecular annotation scarcity and low-quality issues, we conducted a systematic analysis of the ChEMBL database\footnote{https://www.ebi.ac.uk/chembl/}, a well-established life sciences resource encompassing more than 2.5 million compounds and 16,000 targets. As illustrated in~\cref{fig: Data Statistics} (a), we observed the low-quality molecular annotation issue. By removing the abnormal entries, such as null values and duplicate records, we can observe the different distribution between raw molecular activity annotations and denoised annotations. On the other hand, in ~\cref{fig: Data Statistics} (b), we analyzed the IC50 distributions of the top-5 most frequently annotated targets, revealing severe imbalances and wide value ranges across several orders of magnitude.
Hence, in real-world scenarios, existing molecular datasets remain insufficient to support supervised deep learning models. These limitations~\cite{feng2024bioactivity} lead to models that fit the small part of annotated training data but fail to generalize to new molecular structures or properties, an archetypal manifestation of the few-shot problem. 

\noindent \textbf{Few-shot learning on molecular property prediction.} 
As illustrated in~\cref{fig: scenaories comparison}, unlike MPP, few-shot molecular property prediction (FSMPP) alleviates the heavy reliance on large-scale molecular annotations by adopting a small support set with limited supervision and a query set for evaluation, and is formulated as a multi-task learning problem that requires generalization across both molecular structures and property distributions under such constraints.
The primary challenge of FSMPP lies in the risk of overfitting and memorization under limited molecular property annotations, which significantly hampers the generalization ability to new rare chemical properties or novel molecular structures~\cite{guo2021few, wang2021property}.
More specifically, from the perspective of generalization ability, we first identify two essential challenges with the consideration of intrinsic characteristics of molecules: (1) \textit{\textbf{cross-property generalization under distribution shifts}} problem, where different molecular property prediction tasks correspond to distinct structure–property mappings with weak correlations, often differing significantly in label spaces and underlying biochemical mechanisms, thereby inducing severe distribution shifts that hinder effective knowledge transfer;
and (2) the \textit{\textbf{cross-molecule generalization under structural heterogeneity}} problem, where models tend to overfit the structural patterns of a few training molecules and fail to generalize to structurally diverse compounds. 
In light of this, existing FSMPP studies have made various attempts to address these two-fold generalization ability challenges, where external chemical domain knowledge and structural constraints are incorporated into the current FSMPP models for more precise and robust molecular property predictions.

\noindent\textbf{Importance of FSMPP.} Research on FSMPP is becoming increasingly essential for advancing molecular AI systems under real-world constraints. 
Specifically, FSMPP can facilitate early-stage drug discovery by enabling models to predict key pharmacological properties of novel small molecules from only a handful of labeled examples, thereby reducing the need for expensive and time-consuming experimental annotations. For example, FSMPP makes it possible to evaluate the ADMET of candidate compounds even when high-quality labels are scarce, accelerating the prioritization of promising molecules~\cite{niu2024pharmabench}; in therapeutic areas with limited data, such as rare diseases or newly discovered protein targets, FSMPP supports rapid model adaptation to new tasks, allowing predictive insights to generalize across heterogeneous molecular structures and property distributions~\cite{ma2021few,zhang2025improved}. Therefore, addressing FSMPP is crucial for both theoretical advancement and practical applications.

\begin{figure*}[!t]
    \centering
    \resizebox{1\textwidth}{!}{
    \begin{forest} 
    for tree={grow'=0,draw, align=center,anchor=center},
    forked edges,
    [Few-shot Molecular Property \\Prediction Methods \\(\cref{sec: Few-shot}), midbox
        [Data-level~(\cref{sec: Data-level}), midbox
            [Generative Molecule Data\\ Augmentation~(\cref{sec: Generating}), midbox
            [{Meta-MGNN$^{2021}$~\cite{guo2021few}, MTA$^{2023}$~\cite{meng2023meta}, MolFeSCue$^{2024}$~\cite{zhang2024molfescue}}, bigbox]]
            [Implicit Molecule Relation\\ Construction~(\cref{sec: Relations}), midbox
            [{PAR$^{2021}$~\cite{wang2021property},  CPRG$^{2022}$~\cite{yao2022chemical}, Meta-Link$^{2022}$~\cite{cao2022relational},
            IGNTE$^{2023}$~\cite{fifty2023implicit}, \\
            GS-Meta$^{2023}$~\cite{ijcai2023p0526}, KRGTS$^{2025}$~\cite{wang2025knowledge}}, bigbox]]
            [Hybird Methods\\~(\cref{sec: Hybird}), midbox
            [{HSL-RG$^{2023}$~\cite{ju2023few}, PG-DERN$^{2024}$~\cite{zhang2024property}}, bigbox]]
            ],
        [Model-level~(\cref{sec: Model-level}), midbox
            [Molecular Intrinsic Represen-\\tation Learning~(\cref{sec: Intrinsic}), midbox
            [{SMF-GIN$^{2021}$~\cite{jiang2021structure}, Meta-GAT$^{2023}$~\cite{lv2023meta}, PH-Mol$^{2023}$~\cite{zhuang2023prompting}, \\FS-GNNcvTR$^{2023}$~\cite{torres2023convolutional}, FS-GNNTR$^{2023}$~\cite{torres2023few}, APN$^{2024}$~\cite{hou2024attribute}, \\FS-CrossTR$^{2024}$~\cite{torres2024multi}, AttFPGNN-MAML$^{2024}$~\cite{qian2024meta}, \\FS-GCvTR$^{2025}$~\cite{torres2025rethinking}, AdaptMol$^{2025}$~\cite{dai2025adaptmol}}, bigbox]]
            [Molecular Context-aware\\ Learning~(\cref{sec: Context-Aware}), midbox
            [{IterRefLSTM$^{2017}$~\cite{altae2017low}, CAMP$^{2023}$~\cite{fifty2023context}, MHNfs$^{2023}$~\cite{schimunek2023contextenriched}, CRA$^{2024}$~\cite{li2024contextual}, \\ICLPP$^{2024}$~\cite{kaszuba2024context}, MolecularGPT$^{2024}$~\cite{liu2024moleculargpt}, UniMatch$^{2025}$~\cite{li2025unimatch}}, bigbox]]
            ]
        [Learning Paradigm\\~(\cref{sec: Learning Paradigm}), midbox
            [Adapter-based Generalization\\ Strategies~(\cref{sec: Adapter}), midbox[{ATGNN$^{2023}$~\cite{zhang2023adaptive}, EM3P2$^{2023}$~\cite{ham2023evidential}, PACIA$^{2024}$~\cite{wu2024pacia},\\ Pin-Tuning$^{2024}$~\cite{wang2024pin}}, bigbox]]
            [Reformulated Parameter \\Optimization Strategies\\~(\cref{sec: Optimization}), midbox[{ADKF-IFT$^{2023}$~\cite{chenmeta}, QUADRATIC-PROBE$^{2025}$~\cite{formont2025a}}, bigbox]]
            [Other Strategies\\~(\cref{sec: Others}), midbox[{Meta-MolNet$^{2024}$~\cite{lv2024molnet}, AR-APM$^{2024}$~\cite{schimunek2024autoregressive}}, bigbox]]
            ]
        ]
    \end{forest}
    }
    \caption{The systematic taxonomy of existing FSMPP methods.}
    \label{fig: taxonomy}
\end{figure*}
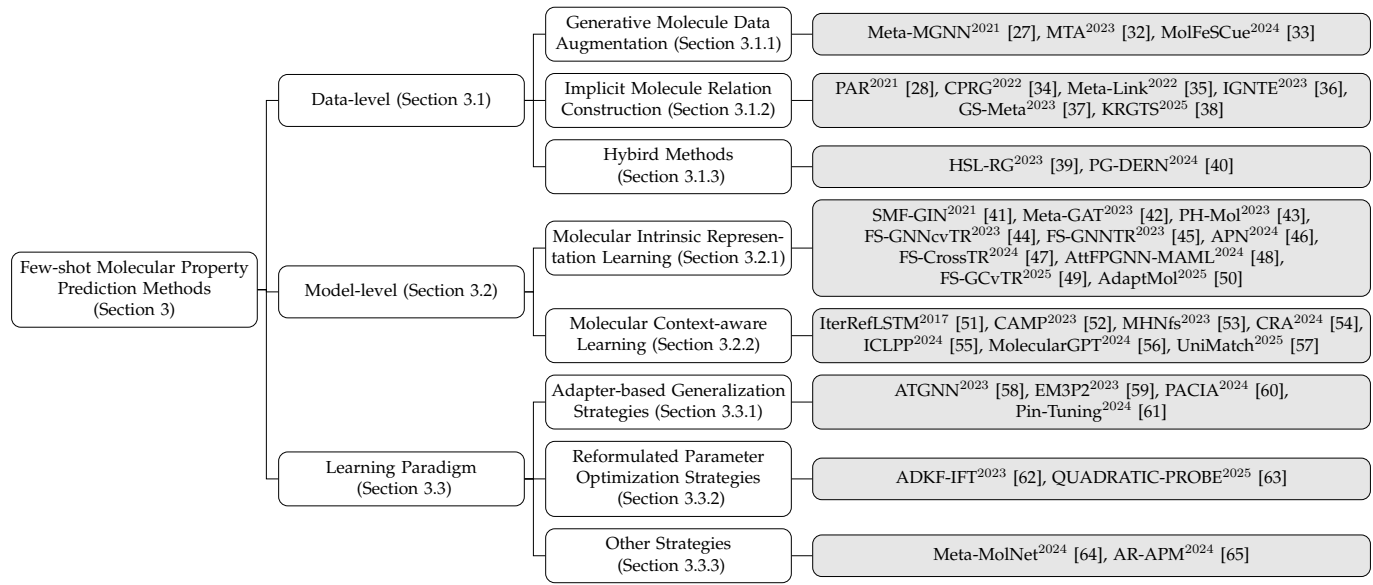
Hence, in this work, to advance the development of FSMPP for practical scenarios,
we conduct a systematic survey, aiming to identify common patterns and fundamental differences across existing few-shot molecular property prediction progress, laying the groundwork for a systematic framework that guides future methodological innovations. Specifically, we categorize existing approaches into data level, model level, and learning paradigm level, along with more fine-grained categories at each level, based on the differences in molecular mining strategies, stages of representation learning, and generalization-oriented optimization mechanisms. Moreover, we compare representative methods, summarize benchmark datasets and evaluation protocols, and identify key future directions to facilitate and inspire continued attention and research on FSMPP.
The main contributions of this survey can be summarized as follows:
\begin{itemize}
    \item We present the first comprehensive survey that systematically reviews the field of FSMPP, offering a timely and valuable reference for both researchers and practitioners in this rapidly evolving domain.
    \item We identify key challenges in FSMPP, including \textit{\textbf{cross-property generalization under distribution shifts}}, i.e., transferring knowledge across weakly correlated tasks with diverse labels and mechanisms, and \textit{\textbf{cross-molecule generalization under structural heterogeneity}}, i.e., the tendency to overfit limited molecular structures. We aim to draw attention to these challenges to inspire further research toward developing more robust and generalizable FSMPP frameworks.
    \item We propose a unified taxonomy that categorizes existing methods from the perspectives of data, models, and learning paradigms, based on a thorough and structured literature review.
    \item We highlight emerging trends and promising directions in FSMPP to guide ongoing efforts and promote sustained advances in this evolving field.
\end{itemize}

\noindent \textbf{Comparison with related surveys.} Despite existing works towards comprehensive reviews on MPP~\cite{li2022deep,tang2023application,kuang2024impact,walters2020applications}, they largely adopt a broad perspective and give little explicit attention to few-shot learning scenarios. In contrast, we argue that the scarcity and often limited quality of molecular property annotations represent more practical challenges—ones that closely align with the current trajectory of AI-assisted molecular research. To the best of our knowledge, this is the first survey specifically dedicated to FSMPP. We aim to provide a systematic and in-depth review of nearly a decade of relevant literature, filling a critical gap in the organization and synthesis of research in this emerging area.

\noindent \textbf{Paper organization.} The remainder of this paper is organized as follows.~\cref{sec: Background} provides background information on FSMPP, including essential concepts and core challenges.~\cref{sec: Few-shot} offers a comprehensive review of state-of-the-art FSMPP methods, along with detailed comparisons and multiperspective discussions.~\cref{sec: Evaluation} introduces commonly used evaluation metrics and benchmark datasets.~\cref{sec: future} highlights open problems and outlines promising future research directions. Finally,~\cref{sec: Conclusion} concludes the survey.

\section{Background}\label{sec: Background}
To provide a common background, Section 2.1 presents an overview of molecular property prediction in a broader context; Section 2.2 formalizes the few-shot learning paradigm; Section 2.3 introduces the few-shot molecular property prediction task; and in Section 2.4, we discuss the core challenges in few-shot molecular property prediction.
\subsection{Molecular Property Prediction}
MPP refers to the task of predicting physicochemical or biological properties of molecules based on their structural representation, playing a fundamental role in domains such as drug discovery~\cite{walters2020applications} and materials science~\cite{hong2020machine}. Depending on the nature of the target properties, MPP tasks are typically formulated as either classification or regression problems. In classification tasks, the goal is to predict discrete labels, including binary classification (e.g., active vs. inactive~\cite{li2020inductive}) and multi-class classification (e.g., assigning a toxicity category~\cite{de2022toxcsm}), whereas regression tasks involve predicting continuous-valued outputs such as solubility, logP, or IC50 values~\cite{park2022comprehensive}, etc.

Generally, given a molecule dataset $\mathcal{D}=\{(x_i, \mathbf{y}_i)\}_{i=1}^{N}$, one can split it as training set $\mathcal{D}_{train}$ and testing set $\mathcal{D}_{test}$ by random splitting or random scaffold splitting. Among them, each molecule $x_i$ is associated with a $\mathbf{y}_i\in \mathbb{R}^{T\times C}$, where $T$ denotes the number of property prediction tasks, and $C$ denotes the number of classes per task (i.e., $C=1$ for regression, $C=2$ for binary classification, and $C>2$ for multi-class classification). The goal of MPP is to learn a function $f^*$ from the training set $\mathcal{D}_{train}$, it can be denoted as:
\begin{equation}
f^*=\arg\underset{f}{\min}~\mathbb{E}_{(x_i,\mathbf{y}_i)\in\mathcal{D}_{\rm train}}~\mathcal{L}(f(x_i),\mathbf{y}_i),
\end{equation}
where $\mathcal{L}$ denotes the task-specific loss function, such as cross-entropy loss for classification or mean squared error (MSE) for regression tasks. Then, it can generalize to the testing set, accurately estimating multiple molecular properties across classification and regression tasks.

Existing MPP methods can be generally categorized into two groups: (1) conventional MPP methods~\cite{khan2016descriptors} and (2) deep learning MPP methods~\cite{li2022deep}. Specifically, conventional MPP methods usually relied on feature engineering, such as molecular descriptors (e.g., molecular weight, logP, TPSA)~\cite{karelson1996quantum,katritzky1995qspr,sun2004universal} and structural fingerprints (e.g., ECFP, MACCS)~\cite{zang2017silico,sturm2018application,huan2015accelerated}, which were designed based on domain knowledge. Then, these features can be fed into conventional machine learning models like the support vector machine (SVM)~\cite{geppert2009ligand} and random forest~\cite{kang2020prediction}, etc. Although interpretable and efficient, these methods are limited by their reliance on predefined rules and inability to capture complex structure–property relationships.

In contrast, deep learning-based approaches have emerged~\cite{walters2020applications}, enabling end-to-end learning of molecular representations. Modern approaches largely differ based on the molecular modality input—namely SMILES strings (1D), molecular graphs (2D), and 3D conformers. Early methods treated SMILES as sequential data and applied language models such as RNNs~\cite{goh2017smiles2vec} and Transformers~\cite{sultan2024transformers} to capture syntactic patterns. In contrast, graph-based approaches represent molecules as 2D topological graphs, allowing graph neural networks~\cite{corso2024graph, jiang2024mix, jiang2025adaptive} to encode connectivity and local environments. Going further, geometric deep learning extends these representations to 3D conformers, incorporating spatial information critical for many biochemical properties~\cite{wang2023automated, moon20233d}. More recently, multi-modal frameworks~\cite{stark20223d, zhang2024pre, wang2024multi} have emerged to jointly leverage 1D, 2D, and 3D views, leading to more robust and generalizable models across diverse prediction tasks. These advances reflect a broader trend that deep MPP models are becoming increasingly chemically grounded, complex, and reliant on large-scale annotated data.

\subsection{Few-shot Learning}
FSL refers to the paradigm of training models under extremely limited supervision. Early FSL studies focus mainly on classification tasks~\cite{wang2020generalizing}. Formally, given a dataset $\mathcal{D}$, it can be split into a support set $\mathcal{S}=\{(x_i, y_i)\}_{i=1}^{C\cdot K}$ and a query set $\mathcal{Q}=\{(x_j, y_j)\}_{j=1}^{N-C\cdot K}$, where $x_i$ denotes an input sample, $y_i\in\{1,\dots, C\}$ is its corresponding label, and $N$ is the number of samples in $\mathcal{D}$. The objective of FSL is to train a model using the limited annotated samples in $\mathcal{S}$ (specifically, the scale of support set far lower than the general supervised training set, i.e., $|\mathcal{D}_{train}|\gg|S|$) and make it generalize well to unseen instances in $\mathcal{Q}$. 
Here, $C$ denotes the number of classes (i.e., $C$-way), and $K$ is the number of samples per class (i.e., $K$-shot), forming the typical $C$-way $K$-shot learning setting. In contrast to few-shot classification, few-shot regression~\cite{SHI2024120010} aims to learn a regression function from only $K$ training samples, where the target variable is continuous. 

Despite differing in objectives, both tasks share the common challenge of learning transferable knowledge from limited supervision. To address this challenge, early efforts in FSL initially adopted the conventional supervised learning paradigm: training models directly on the limited support set $\mathcal{S}$ and evaluating on the query set $\mathcal{Q}$. However, the mismatch between the large parameter space of modern models and the extremely limited supervision often leads to severe overfitting and poor generalization performance. The emergence of transfer learning provided a promising solution~\cite{zhuang2020comprehensive,shin2016deep}. In this setting, a two-stage strategy is employed: the model is first pre-trained on a large-scale source dataset with sufficient supervision, and then fine-tuned using the limited annotated samples from the target task~\cite{sun2019meta,yu2020transmatch}. This approach alleviates overfitting by allowing the model to leverage transferable knowledge learned from related domains. Building upon this idea, meta-learning~\cite{vilalta2002perspective, hospedales2021meta} (i.e., "learning to learn") was introduced to further improve model adaptability across tasks. Rather than training a model for a single task, meta-learning focuses on acquiring an initialization or adaptation strategy that can generalize across a distribution of tasks. These advancements have laid a solid theoretical and practical foundation for FSL and have catalyzed its application in various domain-specific scenarios, including object detection~\cite{fan2020few,sun2021fsce}, image synthesis~\cite{yang2025image} and MPP~\cite{wang2021property,guo2021few}, etc.

\begin{figure*}[!t]
    \centering
    \includegraphics[width=\linewidth]{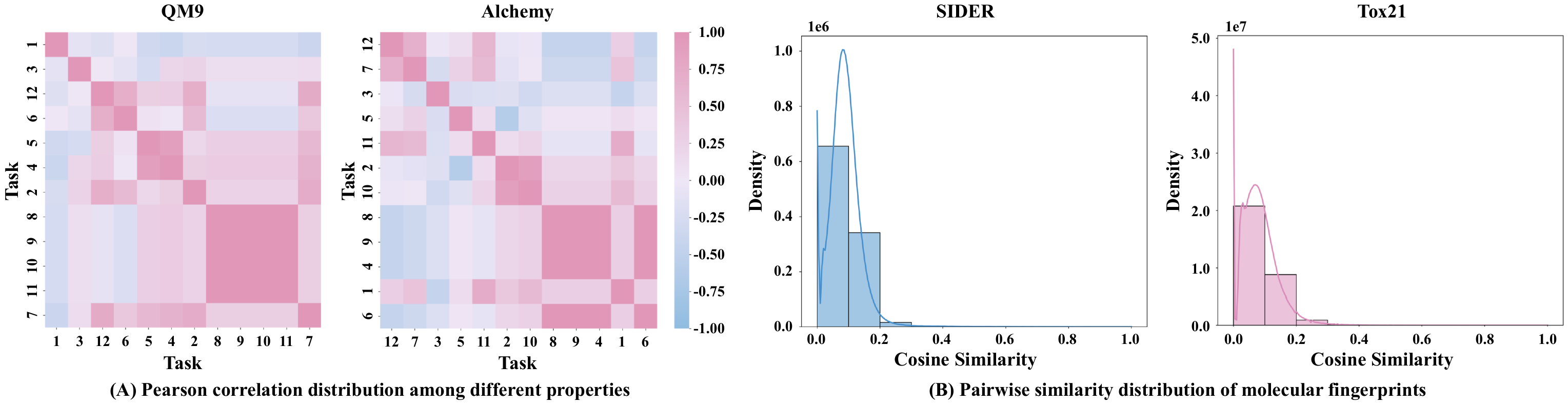}
    \caption{Analysis of properties distribution and molecular structural similarity. (A) Heatmaps showing Pearson correlation coefficients between molecular properties in QM9 and Alchemy. (B) Histogram and density plots of pairwise cosine similarity of molecular fingerprints in SIDER and Tox21. }
    \label{fig: challenge1}
\end{figure*}
\subsection{Few-shot Molecular Property Prediction}
FSMPP aims to accurately predict molecular properties for novel or under-explored properties using limited annotated molecules. 
Given the molecular data annotated with target properties, the task of FSMPP can be formulated as either a classification or regression problem. A key characteristic of FSMPP is its strong alignment with the two-stage meta-learning paradigm widely adopted in general few-shot learning. Specifically, most FSMPP studies are formulated under a multi-task meta-learning framework, where the model is first trained across a distribution of meta-training tasks and then rapidly adapted to new tasks with only a few annotated molecules.

Formally, given a dataset $\mathcal{D}$, it contains a molecule set $X$ and a task set $\mathcal{T}$. Follow the meta-learning setting, the task set can be divided into $\mathcal{T}_{\rm train}$ and $\mathcal{T}_{\rm test}$, corresponding to the meta-training tasks and meta-testing tasks. For each task $\tau\in \mathcal{T}$, it includes a support set $\mathcal{S}_\tau=\{(x_i,y_{i,\tau})\}_{i=1}^{C\cdot K}$ and a query set $Q_\tau=\{(x_i,y_{i,\tau})\}_{i=1}^{N-C\cdot K}$. Similarly, in classification problems, $C$ denotes the number of classes (i.e., $C$-way), and $K$ is the number of samples per class (i.e., $K$-shot), forming the typical $C$-way $K$-shot learning; in regression problems, $C=1$, which means that only $K$ training samples are available. The objective of FSMPP is to learn a model $f^*$ based on meta-training tasks $\mathcal{T}_{\rm train}$ and to achieve generalization in meta-testing tasks $\mathcal{T}_{\rm test}$ with limited supervision:
\begin{equation}
\begin{split} 
    f^* &= \arg\underset{f'}{\min}~\mathbb{E}_{\tau \in\mathcal{T}_{\rm test}}~\mathcal{L}(f',S_{\tau}), \\
    {\rm with}~f'& =\arg\underset{f}{\min}~\mathbb{E}_{\tau\in\mathcal{T}_{\rm train}}~\mathcal{L}(f, S_{\tau}, Q_{\tau}),
\end{split}
\end{equation}
where $\mathcal{L}$ is the loss function of the downstream task.

\subsection{Core Challenges}

In early FSMPP research, efforts typically involved adapting classical FSL algorithms, such as Siamese~\cite{koch2015siamese}, ProtoNet~\cite{snell2017prototypical}, and MAML~\cite{finn2017model}, directly to molecular settings. Although these approaches yielded initial performance gains, their effectiveness was quickly found to be limited in practice~\cite{wang2021property,guo2021few}. This is because FSMPP is not merely a drop-in application of generic FSL methods; instead, it is fundamentally shaped by domain-specific properties such as structural complexity, chemical semantics, and properties heterogeneity. As a result, FSMPP poses two unique and critical challenges as shown in Figure~\ref{fig: challenge1}:

\noindent\textbf{Cross property generalization under distribution shifts.} Unlike conventional FSL scenarios such as image classification, where tasks typically share similar semantic structures, tasks in FSMPP often differ substantially in their underlying chemical mechanisms and structure–property mappings. Each task corresponds to a distinct molecular property (e.g., solubility, toxicity, binding affinity), and the functional relationships between molecular structures and labels can vary significantly. As shown in~\cref{fig: challenge1}~(A), we compute the Pearson correlation coefficients between molecular properties across two representative datasets, QM9~\cite{wu2018moleculenet} and Alchemy~\cite{chen2019alchemy}, and visualize them using heatmaps. The observed correlations are generally weak or even negative, indicating minimal overlap in property-determining features and limited shared structural patterns. Such inter-property distribution shifts—both in structural drivers and label semantics—pose major challenges for learning transferable representations and meta-initializations, thereby complicating generalization across property prediction tasks.

\noindent\textbf{Cross-molecule generalization under structural heterogeneity.} Molecular structures are highly diverse, spanning various functional groups, ring systems, and spatial conformations, leading to severe structural heterogeneity. This poses a fundamental challenge for FSMPP models, which must identify property-relevant patterns from limited support samples. However, the scarcity of data often leads to overfitting to specific molecular scaffolds, ultimately impairing generalization to structurally diverse test molecules. The problem is intensified by the combinatorial complexity of chemical space and the presence of activity cliffs—small structural changes causing large property shifts. Thus, FSMPP models must learn robust, chemically meaningful representations that generalize across structurally dissimilar molecules. To quantify this diversity, we compute pairwise cosine similarity of molecular fingerprints in the SIDER and Tox21~\cite{wu2018moleculenet}. As shown in~\cref{fig: challenge1}~(B), most molecular pairs have similarities below 0.2, highlighting substantial intra-task structure heterogeneity and the difficulty of learning transferable structure–property patterns.

Together, these challenges highlight the inadequacy of directly transplanting existing FSL methods into molecular domains. They motivate the development of FSMPP-specific strategies, including cross-property invariant representations, sample-efficient structure encoders, and chemically aware meta-learning paradigms.


\section{Few-shot Molecular Property Prediction Methods}\label{sec: Few-shot}
To provide a comprehensive understanding of how FSMPP methods tackle data scarcity and generalization challenges, we categorize existing approaches into three interconnected levels: data-level, model-level, and learning paradigm. This taxonomy reflects how current methods extract and leverage prior knowledge from different angles. Specifically, the data level focuses on generating novel samples or mining implicit structural and relational information from limited data to enhance generalization across properties and molecules; the model level concentrates on learning chemically meaningful and transferable representations that can adapt to unseen properties and structures; and the learning paradigm level concerns how models are trained and fine-tuned under few-shot constraints, often involving multi-stage processes such as task adaptation and meta-training. These three levels represent essential components of an FSMPP framework and work collaboratively to improve performance in FSMPP.

\begin{figure*}[!t]
    \centering
    \includegraphics[width=\linewidth]{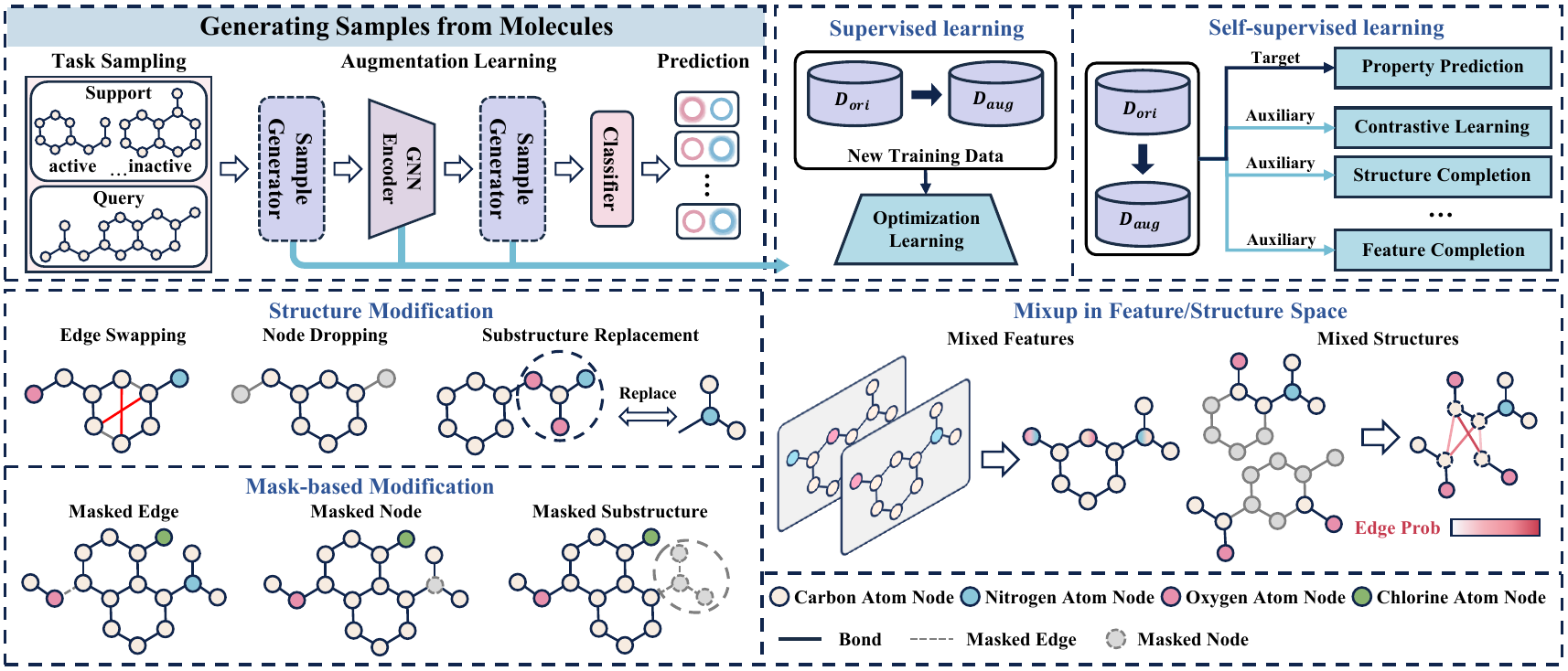}
    \caption{Solving FSMPP problems by generative molecule data augmentation. This framework shows both how to generate new samples and how to effectively leverage them. Sample generation strategies include: (1) structure modification, such as bonds swapping, atoms dropping, and substructure replacement, etc.; (2) mixup in feature or structure space, which blends molecular features or graph structures; and (3) mask-based modification, where atoms, bonds, or substructures are masked to construct auxiliary learning signals. The generated samples are utilized through two learning paradigms: (1) a supervised route, where augmented data is directly added to the training set to alleviate data scarcity; and (2) a self-supervised route, where tasks such as molecule completion or contrastive learning are introduced to enhance models.}
    \label{fig: Generating Samples}
\end{figure*}


\subsection{Data-level Methods}\label{sec: Data-level}
To address the fundamental challenge of data scarcity and low quality in FSMPP,  meanwhile, breaking the limitation in cross-property/structure adaptation, data-level methods provide the most direct and intuitive solutions. These methods mainly fall into three directions: 
(1) \textit{Generative molecule data augmentation}, i.e., expanding molecular samples to simulate richer task distributions and increase training diversity in a generative way;
(2) \textit{Implicit molecule relation construction}, mining latent information from the limited available data, such as exploiting implicit relations among molecules or properties, to compensate for the lack of supervision;
(3) \textit{Hybrid methods} that combine both sample-level augmentation and relational signal modeling to enhance data quantity and enrich the supervised information jointly.

\subsubsection{Generative Molecule Data Augmentation}\label{sec: Generating}
To address the data scarcity problem, an intuitive and widely adopted strategy is to generate new samples from existing molecules. Existing methods usually introduce different kinds of transformations to accessible data to improve the data diversity. Furthermore, one can directly use the enriched dataset to train a more generalizable model or introduce some self-supervised learning tasks to enhance the model. Although conceptually straightforward, such data-level methods have proven remarkably powerful and have emerged as a vibrant subfield. Fundamentally, these methods revolve around two key questions: (1) How to design effective strategies for generating new samples, and (2) How to leverage the generated samples to improve model generalization ability.

\noindent\textbf{How to generate new samples.} Most generation strategies are designed to introduce structural variations while preserving the underlying chemical validity, with the goal of simulating plausible molecules that could help regularize the learning process. As shown in~\cref{fig: Generating Samples}, the following categories summarize the mainstream approaches:
\begin{itemize}[leftmargin=*, itemsep=0.3em]
    \item Structure Modification: This strategy perturbs the molecular structure by modifying atoms, bonds, or substructures to generate chemically valid variants. Typical operations include atom or bond addition, deletion, substitution, bond order alteration, and the insertion or replacement of functional groups, ring systems, or scaffolds. These modifications~\cite{ju2023few} introduce localized or global structural diversity while preserving overall molecular plausibility, enabling the model to learn more robust and generalizable representations across varied chemical contexts.
    \item Mixup in Feature/Structure Space: Inspired by the mixup strategy in image and language domains, this approach interpolates either the representation or structure of two molecules. It encourages smoother decision boundaries and enhances the model's generalization ability. For example, Meng et al.~\cite{meng2023meta} propose Motif-based Task Augmentation (MTA), which mixes molecular embeddings with those of retrieved motifs to generate augmented support/query samples. To ensure semantic consistency, MTA retrieves molecule-motif pairs based on task-specific representation similarity, thereby aligning augmented samples with underlying property distributions.
    \item Mask-based Modification: This technique involves masking specific atoms, bonds, or substructures within a molecule and formulating a learning task that requires the model to predict the masked elements. It serves both pre-training and data augmentation purposes, enabling the model to learn contextualized molecular representations through reconstruction-based objectives. This strategy is often integrated with contrastive learning frameworks or denoising autoencoder paradigms. For example, recent works~\cite{guo2021few, zhang2024molfescue} adopt atom and bond masking as a core augmentation technique, incorporating auxiliary tasks such as atom type prediction and bond reconstruction to provide richer supervision and improve performance under few-shot scenarios.
\end{itemize}

\noindent\textbf{How to leverage the generated samples.} Once generated, these synthetic samples can be used in two primary ways: 1) Supervised learning: A simple application of generated molecular samples is to directly incorporate them into the supervised training set. By augmenting the empirical data distribution, this approach increases the diversity and density of training samples, thereby mitigating overfitting and enhancing generalization. This strategy is particularly effective in scenarios with severely limited support sets, where the inclusion of augmented molecules can enrich the decision boundaries and provide more informative training signals for the model. Such as MTA~\cite{meng2023meta} simultaneously augment the support set and the query set. This design aligns with the two-stage optimization paradigm of FSMPP, offering additional knowledge for outer-loop model updates while enhancing the mutual information between the support and query sets. Consequently, it effectively mitigates the problems posed by cross-property and cross-structure generalization in FSMPP. 2) Self-supervised learning: An alternative strategy involves incorporating generated or masked molecular variants into self-supervised learning frameworks, where they serve as inputs for auxiliary tasks. These tasks, such as contrastive learning~\cite{zhang2024molfescue}, masked component prediction~\cite{guo2021few,ju2023few}, and other reconstruction-based objectives, encourage the model to capture structural semantics and contextual dependencies beyond target property supervision. In doing so, the model learns more robust and generalizable molecular representations, which are particularly beneficial in low-resource settings.

\begin{figure*}[!t]
    \centering
    \includegraphics[width=\linewidth]{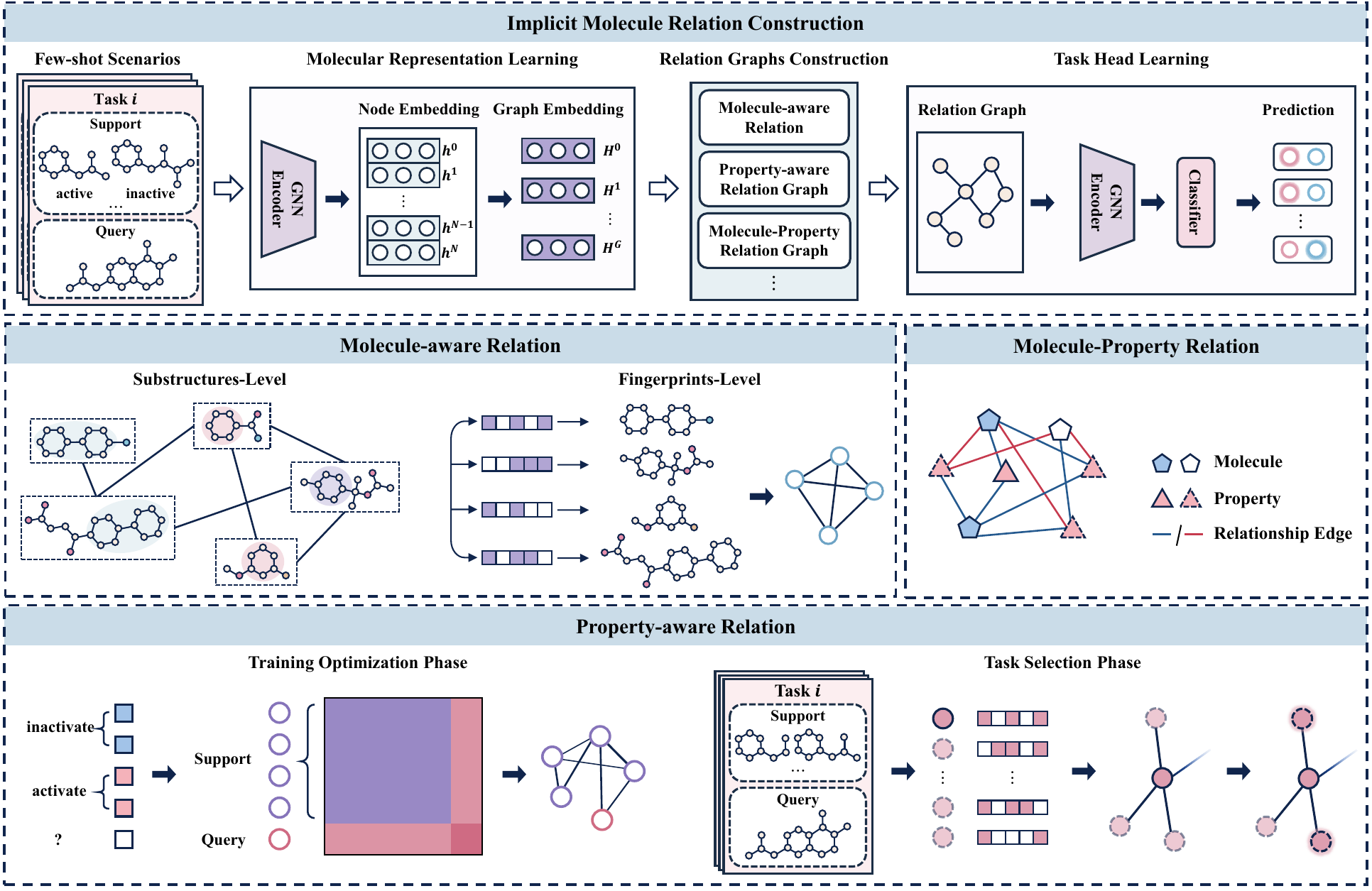}
    \caption{Solving FSMPP problems by implicit molecule relation construction. This framework extends the two-stage FSMPP pipeline by integrating relation graph learning mechanisms to capture the many-to-many associations between molecules and properties. Specifically, molecular embeddings are first extracted using a molecular encoder, followed by the construction of relation graphs. These graphs are then utilized by relational learning modules to enhance molecular representations or guide the meta-training process, thereby improving generalization in few-shot scenarios.}
    \label{fig: implicit relations}
\end{figure*}

\subsubsection{Implicit Molecule Relation Construction}\label{sec: Relations}
Another important branch of data-level approaches focuses on mining and modeling the latent relational structures underlying the associations between molecules and their target properties. Molecular property prediction is inherently governed by complex and interdependent relationships—where a single molecule may exhibit multiple properties, and structurally diverse molecules may share common properties due to substructure-level similarities. Additionally, certain properties often co-occur across molecules, reflecting latent functional or mechanistic dependencies. Capturing such multifaceted relations—including molecule-molecule relation, molecule-property relation, and property-property relation—is critical for enabling robust generalization in few-shot settings. To this end, recent methods have proposed relation-based frameworks that construct task-specific or domain-informed relational graphs based on molecular representations or external priors (e.g., fingerprints). These methods typically follow a two-stage meta-learning paradigm: molecular encoders, such as graph neural networks~\cite{xu2018how, Song2020Communicative}, first generate representations for support and query samples, which are then used to construct relation graphs encoding structural or semantic dependencies. These graphs are further processed by relation learning modules to enhance molecular representations for property prediction. Central to this pipeline is the design of the relation graph itself, which underpins the model’s ability to transfer knowledge from limited data. Existing approaches in this direction can be broadly categorized into three types, based on the nature of the constructed relations and their modeling objectives.

\noindent \textbf{Molecule-aware relation graph.} These methods construct graphs where nodes represent individual molecules and edges encode task-relevant interactions~\cite{wang2021property} or structural similarity~\cite{ju2023few}. Similarity metrics may include Tanimoto similarity based on fingerprints and representation similarity, etc. The core intuition is that molecules with similar structures or shared substructures tend to exhibit similar properties. A representative and classical method in this line of research is the Property-Aware Relation Network (PAR) proposed by Wang et al.~\cite{wang2021property}. PAR is among the first to explicitly model the variability of relevant substructures and molecule-molecule relationships across different molecular properties. It constructs a molecule-level relation graph using property-aware embeddings, allowing the model to capture task-specific structural dependencies and improve generalization in few-shot molecular property prediction.

\noindent \textbf{Property-aware relation graph.} Moreover, in FSMPP settings, each task typically corresponds to a distinct molecular property. Recent studies have begun to explore latent similarities or dependencies among tasks, aiming to identify and leverage underlying chemical or functional relationships between them. However, the connections between molecular properties are highly domain-specific and often involve deep expert knowledge. Existing methods therefore advocate capturing these relationships from property-driven molecular similarities rather than relying on explicit expert annotations. One representative approach is CPRG~\cite{yao2022chemical}, which introduces a chemical property relation modeling framework to construct a property-aware relation graph. By identifying positively correlated properties through Spearman correlation on property-aware molecular representations, CPRG guides meta-training with related tasks, enabling more effective knowledge transfer and improved few-shot prediction performance.

\noindent \textbf{Molecular-property relation graph.} Instead of focusing solely on molecule-level or property-level connections, MetaLink~\cite{cao2022relational} argues that the known properties of molecules are helpful to target properties prediction and first develops the molecule-property relation graph in FSMPP. Among them, nodes denote molecules and their associated properties, and each edge denotes the label of a molecule on a particular property. Such a relational modeling framework can effectively model knowledge transfer from auxiliary property labels to the target properties. Moreover, IGNTE~\cite{fifty2023implicit} models the molecule-protein relations and leverages large-scale docking simulation data for multi-task pretraining of molecular embeddings, which are then adapted to downstream FSMPP tasks using meta-learning strategies. 

Overall, modeling implicit molecular-property relations enables FSMPP models to move beyond isolated sample representations and leverage the inherent prior knowledge embedded in molecular datasets. Specifically, molecular-level relation graphs help capture structural semantics and support generalization to structurally diverse compounds. Meanwhile, task-level or property-level relation graphs facilitate the modeling of cross-property dependencies, contributing to improved transferability across molecular properties. However, relying on a single type of relational information may constrain the scope of knowledge transfer and limit the model’s ability to address the multifaceted challenges of FSMPP. To address this, GS-Meta~\cite{ijcai2023p0526} extends prior approaches by enriching the molecule-property relation graph with additional molecule-molecule connections. It introduces a meta-training subgraph sampling strategy centered on target properties, guided by a contrastive learning objective. This unified relational structure allows for the joint modeling of intra-molecular and cross-property interactions within a single framework. Building on this idea, KRGTS~\cite{wang2025knowledge} incorporates a reinforcement learning-based auxiliary property sampling mechanism, which dynamically selects auxiliary properties based on their semantic relevance to the target property. In addition, it introduces a substructure-driven strategy to refine molecule-molecule relations at a finer granularity, aiming to improve the precision of relational modeling. These developments reflect a broader trend toward integrating multi-level relational signals to better capture the complex many-to-many mappings between molecules and their properties in FSMPP.

\begin{table*}[!t]
  \centering
  \caption{Characteristics for FSMPP methods focusing on the data-level.}
    \resizebox{0.9\linewidth}{!}{
    \begin{tabular}{cccccc}
    \toprule[1.5pt]
    Category & Method & Year  & Venue & Generation Strategy & Relation Type \\
    \midrule
\multirow{3}[1]{*}{Generative Molecule Data Augmentation} & Meta-MGNN\cite{guo2021few} & 2021  & WWW   & Atom/Bond Mask & - \\
     & MTA~\cite{meng2023meta}   & 2023  & SDM   & Feature Mixup & - \\
     & MolFeSCue~\cite{zhang2024molfescue} & 2024  & Bioinformatics & Atom/Bond Mask & - \\
    \midrule
    \multirow{9}[1]{*}{Implicit Molecule Relation Construction} & PAR~\cite{wang2021property}   & 2021  & NeurIPS & -     & Molecule-Molecule \\
     & CPRG~\cite{yao2022chemical} & 2022  & IJCNN & -     & Property-Property \\
     & Meta-Link~\cite{cao2022relational} & 2022  & ICLR  & -     & Molecule-Property \\
     & IGNTE~\cite{fifty2023implicit} & 2023  & NeurIPS-MLSBW & -     & Property-Property \\
    \cmidrule{2-6}
     & \multirow{2}[0]{*}{GS-Meta~\cite{ijcai2023p0526}} & \multirow{2}[0]{*}{2023} & \multirow{2}[0]{*}{IJCAI} & \multirow{2}[0]{*}{-} & Molecule-Molecule \\
     &       &       &       &       & Molecule-Property \\
    \cmidrule{2-6}
     & \multirow{3}[1]{*}{KRGTS~\cite{wang2025knowledge}} & \multirow{3}[1]{*}{2025} & \multirow{3}[1]{*}{Information Sciences} & \multirow{3}[1]{*}{-} & Molecule-Molecule \\
     &       &       &       &       & Molecule-Property \\
     &       &       &       &       & Property-Property \\
    \midrule
    \multirow{2}[1]{*}{Hybrid Methods} & HSL-RG~\cite{ju2023few} & 2023  & Neural Network & \multicolumn{1}{l}{Substructure Modification} & Molecule-Molecule \\
     & PG-DERN~\cite{zhang2024property} & 2024  & IEEE JBHI & Feature Mixup & Molecule-Molecule \\
    \bottomrule[1.5pt]
    \end{tabular}%
    }
  \label{tab: data-level}%
\end{table*}%

\subsubsection{Hybird Methods}\label{sec: Hybird}
While generating new molecules and modeling implicit relationships have each demonstrated notable advantages independently, recent studies have increasingly explored hybrid methods that integrate both strategies within a unified framework. These approaches aim to simultaneously expand the molecular sample space and uncover latent relational structures, thereby enriching the supervision signal from both data and knowledge perspectives. Representative methods include HSL-RG\cite{ju2023few}, which introduces global molecule-molecule relation graphs constructed via graph kernel similarities and develops the replacements a local self-supervised structure perturbation module to capture both inter-molecular dependencies and transformation-invariant features. PG-DERN\cite{zhang2024property} employs a dual-view encoder to learn fine- and coarse-grained molecular representations, constructs similarity-based molecular relation graphs, and introduces a representation mixup strategy guided by auxiliary property information, facilitating improved generalization to novel property prediction tasks.

\subsubsection{Discussion and Summary}
To address the challenge of data scarcity in FSMPP, meanwhile, breaking the limitation in cross-property/structure generalization, two primary data-level strategies have been explored: generative molecule data augmentation and implicit molecule relation construction. The former focuses on augmenting the training set by generating molecular variants, either directly or through self-supervised tasks, to help models learn invariant features and generalize to structurally diverse compounds. The latter emphasizes mining the latent knowledge within molecular datasets by modeling molecule-molecule, molecule-property, and property-property relationships, thereby enabling cross-structure and cross-property generalization. These two approaches are conceptually complementary since one expands the data space, while the other leverages intrinsic prior knowledge.

Despite their strengths, each approach faces inherent challenges. For sample generation, a major concern is label consistency, as small structural changes may result in significant shifts in molecular properties, potentially introducing noise. This motivates the use of mask-based strategies, which offer greater control and biological plausibility. In relational modeling, the key difficulty lies in accurately capturing and utilizing meaningful relationships, which depend heavily on representation quality and graph construction. Recent hybrid methods attempt to integrate both directions, combining structural augmentation with relational reasoning to achieve better generalization. Overall, the trend in FSMPP is shifting toward multi-source information integration. Representative methods are summarized in~\cref{tab: data-level}.

\subsection{Model-level Methods}\label{sec: Model-level}
While data-level methods focus on alleviating sample scarcity, model-level methods aim to enhance the model’s capacity to learn informative representations under limited supervision. Given the complexity and heterogeneity of molecular structures, it is essential to design architectures that can effectively capture intrinsic structural features and adapt to varying molecular contexts. Therefore, model-level approaches in FSMPP are typically grounded in two directions: (1) Molecular intrinsic representation learning, i.e., learning robust intrinsic representations of molecular structures themselves; and (2) Molecular context-aware learning, i.e., incorporating property-specific or molecule-specific context to improve adaptability and generalization.

\begin{figure}[!t]
    \centering
    \includegraphics[width=\linewidth]{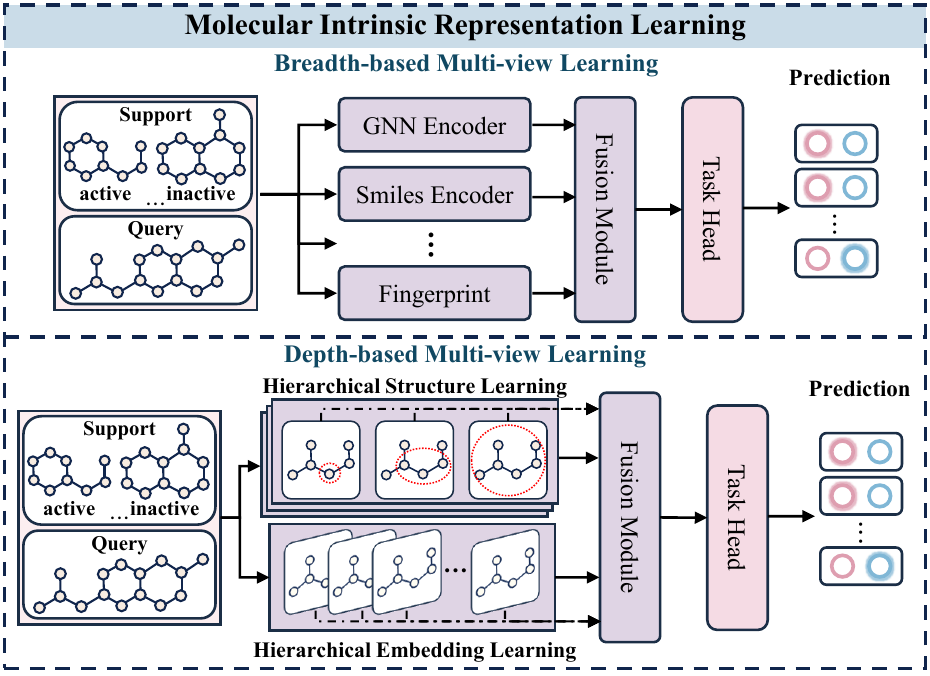}
    \caption{Solving FSMPP problems by molecular intrinsic representation learning. These methods focus on optimizing molecular representations from themselves to address FSMPP challenges. Breadth-based multi-view learning integrates different molecular modalities (e.g., GNN, SMILES, fingerprint), while depth-based multi-view learning aggregates hierarchical information from molecular structures and embeddings to enrich representations.}
    \label{fig: model-level1}
\end{figure}

\subsubsection{Molecular Intrinsic Representation Learning}\label{sec: Intrinsic}
To address the challenges posed by the heterogeneity of molecular structures under few-shot scenarios, molecular intrinsic representation learning has focused on advancing molecular representation learning techniques. These methods~\cite{qian2024meta, hou2024attribute, jiang2021structure, torres2023convolutional, lv2023meta, torres2023few, zhuang2023prompting, torres2024multi, torres2025rethinking, dai2025adaptmol} typically operate at the level of individual molecules, seeking to extract rich and informative features that reflect both fine-grained structural details and broader molecular characteristics. A common foundation across this line of work is the integration of multiple complementary views of each molecule, enabling the representation to capture diverse aspects of molecular semantics within a unified framework, thereby improving model performance in few-shot prediction tasks. In this section, we review representative methods that follow this design paradigm. As shown in~\cref{fig: model-level1}, based on the underlying strategy used to extract and integrate multi-view information, these methods can be categorized into breadth-based multi-view molecular representation learning and depth-based multi-view molecular representation learning. In particular, to better characterize and compare different techniques, we summarize the key features of molecular intrinsic representation learning methods in~\cref{tab: model-level}.


\noindent \textbf{Breadth-based multi-view molecular representation learning.} Breadth-based multi-view molecular representation learning typically enhances representational expressiveness by integrating molecular representations derived from different views. Instead of relying solely on a single molecular encoding method, these methods incorporate multiple complementary views, such as molecular fingerprints, thereby enriching the representational space and improving generalization in few-shot settings. For example, AttFPGNN-MAML~\cite{qian2024meta} introduces three complementary molecular fingerprints - MACCS, Pharmacophore ErG, and PubChem, combining graph neural networks (GNN) and hybrid molecular fingerprints to generate a more comprehensive molecular representation. It utilizes an instance attention mechanism to capture similarities and differences among molecules, generating task-specific molecular representations. Similarly, APN~\cite{hou2024attribute} first extracts molecular attributes from various molecular fingerprints and self-supervised learning methods, then designs an attribute-guided dual-channel attention module to learn the relationship between molecular graphs and attributes, and makes final predictions through a prototype network. This approach generates more informative molecular representations by learning the relationship between molecular graphs and attributes through the molecular attribute extractor and the attribute-guided dual-channel attention module. 
\begin{table*}[!t]
  \centering
  \caption{Characteristics for FSMPP methods focusing on the molecular intrinsic representation learning.}
    \resizebox{1\linewidth}{!}{
    \begin{tabular}{ccccc}
    \toprule[1.5pt]
    Category & Method & Year  & Venue & Characteristic \\
    \midrule
    \multirow{2}{*}{\textbf{Breadth-based Multi-view}} 
      & AttFPGNN-MAML \cite{qian2024meta} & 2024 & ACS\ OMEGA & Combining molecular fingerprints \\
      & APN \cite{hou2024attribute}      & 2024 & BIB      & Combining molecular fingerprints \\
    \midrule
    \multirow{5}{*}{\textbf{Depth-based Multi-view}} 
      & SMF-GIN \cite{jiang2021structure}          & 2021 & AI Open         & Hierarchical molecular structure learning \\
      & Meta-GAT \cite{lv2023meta}                & 2023 & TNNLS           & Hierarchical molecular structure learning \\
      & FS-GNNTR \cite{torres2023few}             & 2023 & ESWA            & Hierarchical molecular embedding learning \\
      & FS-GNNCvTR \cite{torres2023convolutional} & 2023 & ESANN           & Hierarchical molecular embedding learning \\
      & FS-GCvTR \cite{torres2025rethinking}      & 2025 & Pattern Recognition & Hierarchical molecular embedding learning \\
    \midrule
    \multirow{2}{*}{\textbf{Breadth + Depth-based}} 
      & PH-Mol \cite{zhuang2023prompting}         & 2023 & MedAI           & Combining task description \& Hierarchical molecular structure learning \\
      & AdaptMol \cite{dai2025adaptmol}           & 2025 & arXiv           & Combining SMILES \& Hierarchical molecular structure learning \\
    \bottomrule[1.5pt]
  \end{tabular}
    }
  \label{tab: model-level}%
\end{table*}%

\noindent \textbf{Depth-based multi-view molecular representation learning.} Depth-based multi-view molecular representation learning focuses on modeling the hierarchical structural information within individual molecules. Unlike breadth-based methods that integrate heterogeneous external representations, these approaches aim to capture internal structural dependencies by organizing molecular information across multiple levels of granularity, such as atoms, bonds, substructures, and global molecular topology. Some methods achieve this by explicitly encoding topological components of molecules from different structural levels. For example, SMF-GIN~\cite{jiang2021structure} introduces a local structure attention module that individually encodes key subgraphs and assigns different weights based on an attention mechanism. This allows for the automatic learning of weight distributions across different layers and substructures, ultimately merging with the global representation to effectively combine multi-scale and multi-level molecular representations. Similarly, Meta-GAT~\cite{lv2023meta} aggregates feature information from atoms, bonds, and functional groups through a triple attention mechanism to capture the local influences of atom groups. It enhances the understanding of complex molecular topologies through multi-step iterations and employs a bidirectional GRU combined with an attention mechanism to obtain high-quality global molecular representations. Other approaches, rather than decomposing molecular structures directly, construct multiple views by transforming or segmenting the learned molecular representations. For instance, FS-GNNTR~\cite{torres2023few} serializes molecular graph embeddings and applies positional encoding followed by self-attention and feedforward layers, enabling the model to capture expressive contextual dependencies. FS-GNNCvTR~\cite{torres2023convolutional} further enhances local structural modeling by tokenizing molecular graph embeddings into patches and applying convolutional operations to generate localized token sequences. FS-CrossTR~\cite{torres2024multi} introduces a dual-branch multi-scale Transformer that processes graph embeddings of different resolutions in parallel, using cross-attention to align and integrate structural information across scales. More recently, FS-GCvTR~\cite{torres2025rethinking} segments graph embeddings into spatial patches and employs a convolution Transformer to jointly learn local and global dependencies, further advancing fine-grained molecular representation learning.

In addition, some recent methods integrate both depth-based and breadth-based multi-view learning mechanisms to further enhance the expressiveness and generalizability of molecular representations. These hybrid approaches combine hierarchical structural modeling with multimodal feature integration. For example, PH-Mol~\cite{zhuang2023prompting} employs a hierarchical graph neural network to capture molecular structures at both atomic and group levels, and further integrates property-descriptive prompts to guide molecular representation learning, enabling property-aware and structurally rich predictions under few-shot settings. More recently, AdaptMol~\cite{dai2025adaptmol} integrates an adaptive multi-view learning mechanism by combining information from both SMILES sequences and molecular graphs to learn more informative molecular representations. It introduces an adaptive multi-level attention module that dynamically integrates local and global molecular features.

\subsubsection{Molecular Context-aware Learning}\label{sec: Context-Aware}
Molecular context-aware learning focuses on enhancing molecular representations by modeling the contextual dependencies within and across molecules. Instead of relying solely on static structural features, these methods dynamically adjust the feature extraction process based on the property-specific context or molecular environment, thereby improving the generalization ability and adaptability of models under few-shot settings. Representative methods~\cite{altae2017low, fifty2023context, schimunek2023contextenriched, li2024contextual, li2025unimatch} typically adopt attention-based architectures, memory-enhanced modules, or large language models to capture fine-grained contextual cues and enable robust knowledge transfer across molecular structures and properties.

Among the earlier efforts, IterRefLSTM\cite{altae2017low} adopts a GCN-LSTM framework that embeds both query and support molecules into a shared vector space, and iteratively refines the query embedding via an attention mechanism that weighs the most relevant support molecules. This facilitates effective knowledge transfer in sparse-data scenarios by allowing the model to dynamically focus on informative examples. Building on joint modeling of molecular structures and label semantics, CAMP~\cite{fifty2023context} constructs contextual demonstration sequences by concatenating molecular and label embeddings, which are fed into a Transformer encoder to learn adaptive representations aligned with task-specific classification boundaries. Such label-aware alignment improves the ability to generalize across heterogeneous molecular properties. To further enrich contextual understanding, MHNfs\cite{schimunek2023contextenriched} introduces a multi-module architecture that retrieves relevant molecules from a large-scale memory pool using the Modern Hopfield Network, exchanges information via cross-attention between query and support molecules, and finally estimates similarity to drive few-shot prediction. This modular structure enables effective representation enrichment, even in highly imbalanced or diverse molecular distributions. Extending this line of work, CRA~\cite{li2024contextual} introduces a context-aware representation anchor mechanism to address the selection bias that often arises in support-query pairings. It constructs a query-aware support context and incorporates cross-properties semantic memory to learn context-sensitive prototypes. Pushing the boundary of cross-task generalization, UniMatch\cite{li2025unimatch} unifies atom-level and task-level matching via a dual-channel strategy: the explicit branch uses hierarchical attention to match molecular substructures, while the implicit branch employs meta-learning to align task-level parameters. This unified perspective equips the model with strong generalization ability across both structural and task domains.

Beyond these neural architecture-based approaches, recent advances explore the use of large language models to capture broader molecular context and support cross-domain generalization. For instance, ICLPP~\cite{kaszuba2024context} explores in-context learning  for molecular property prediction by coupling a geometry-aware graph neural network with a GPT-2 language model. The model is trained to infer structure–property relationships from sequences of molecules that share common substructures, enabling GPT-2 to adaptively learn from atomic-level geometric features. Rather than relying on extensive fine-tuning, it leverages contextual molecular prompts to generalize effectively to out-of-distribution molecules, demonstrating significant performance gains on the QM9 benchmark. MolecularGPT~\cite{liu2024moleculargpt} explores molecular context-aware learning in FSMPP by fine-tuning a large language model on a molecular instruction set spanning biology, chemistry, and quantum mechanics. It enhances few-shot reasoning by retrieving structurally similar molecules as in-context demonstrations, enabling the model to leverage structural context for improved generalization across low-resource molecular property prediction tasks.

\subsubsection{Discussion and Summary}
Model-level methods in FSMPP reflect two complementary yet fundamentally distinct strategies for improving generalization under limited supervision: Molecular Intrinsic Representation Learning and Molecular Context-Aware Learning. The former focuses on capturing the internal structural richness of molecules by leveraging multi-view integration techniques. Breadth-based approaches aggregate diverse external descriptors (e.g., fingerprints, attributes, SMILES), while depth-based methods emphasize hierarchical structural granularity, organizing atoms, bonds, and substructures into multi-level topological representations. These strategies aim to enhance molecular expressivity and within-task performance by constructing structurally comprehensive embeddings for each molecule.

In contrast, molecular context-aware learning shifts the focus from individual molecular structures to inter-molecular relationships and property-level context. These methods dynamically adapt feature extraction and representation learning based on the support-query composition of few-shot tasks, enabling better knowledge transfer across tasks and structural variations. By utilizing attention mechanisms, memory-augmented modules, or LLM-based in-context learning, such approaches demonstrate strong adaptability in out-of-distribution settings and cross-task scenarios. However, while intrinsic representation learning provides structural completeness, it may struggle with generalization; context-aware methods enhance adaptability but depend heavily on the quality of contextual signals. Therefore, future directions may lie in hybridizing both paradigms, jointly leveraging structural fidelity and contextual flexibility to enhance performance across diverse molecular prediction tasks.

\subsection{Learning Paradigm}\label{sec: Learning Paradigm}
Beyond data and model design, another key to addressing the challenge of FSMPP lies in the learning paradigm itself: how the model is optimized and adapted to achieve generalization from a few examples. Learning paradigm-level methods in FSMPP focus on developing principled optimization strategies that improve the model’s adaptability, generalization, and robustness. These approaches include: (1) Adapter-based generalization strategies, i.e., utilizing adapter modules to introduce flexible and task-specific adaptation mechanisms; (2) Reformulated parameter optimization strategies, i.e., redefining the optimization objectives or learning algorithms to better fit the few-shot setting; and (3) Other strategies, i.e., mostly exploring alternative training routines or frameworks.

\begin{table*}[!t]
    \caption{Comparison of adapter-based generalization strategies.}
    \centering
    \renewcommand{\arraystretch}{1.1}
    \resizebox{\linewidth}{!}{
    \begin{tabular}{ccccc}
    \toprule[1.5pt]
    Method & Adaptation operate position & Parameter optimization strategy & Adaptation Granularity  \\
    \midrule
    ATGNN~\cite{zhang2023adaptive} & Molecular encoder & Parameter mixup & Full model parameters \\
    EM3P2~\cite{ham2023evidential} & Prediction head & Output regularization & Prediction layer \\
    PACIA~\cite{wu2024pacia} & Molecular encoder\& Prediction head & Local updates with hypernetwork parameters & Adapter Layer  \\
    Pin-Tuning~\cite{wang2024pin} & Molecular encoder\& prediction Head & Local updates with weight regularization & Adapter Layer  \\
    \bottomrule[1.5pt]
    \end{tabular}
    }
    \label{tab: Adapter}
\end{table*}

\subsubsection{Adapter-based Generalization Strategies}\label{sec: Adapter}
Due to the extremely limited supervision in FSMPP, full fine-tuning of large neural architectures often leads to overfitting or inefficiency. To address this, adapter-based approaches introduce lightweight, trainable mechanisms into FSMPP framework, enabling task-aware adaptation with minimal parameter updates. This design allows the model to retain general molecular knowledge while flexibly adapting to new tasks under few-shot constraints. Four representative methods, such as ATGNN~\cite{zhang2023adaptive}, EM3P2~\cite{ham2023evidential}, PACIA~\cite{wu2024pacia}, and Pin-Tuning~\cite{wang2024pin}, differ in their architectural choices, but they share two main technical directions: (1) Where adaptation operates in the framework, to enable lightweight task-aware adaptation, and (2) How parameter optimization is performed, to support fast and efficient learning. For the first, adaptation mechanisms are typically inserted into the molecular encoder and prediction head to modulate intermediate representations and outputs. For the second, optimization strategies include meta-learned parameter generation, interpolation between pre-trained and fine-tuned weights, local tuning, and output reparameterization. These design choices collectively shape the efficiency and generalization ability of FSMPP models. A summary of representative methods is shown in~\cref{tab: Adapter}.

\noindent\textbf{Where adaptation operates in the framework.} In FSMPP, it is important to adapt the model architecture according to the support task to handle distribution shifts. Current advances in FSMPP primarily focus on enhancing two key components of the molecular prediction architecture: the molecular encoder, which is responsible for learning effective structural representations of molecules, and the prediction head, which translates these representations into property-specific predictions. To enable task-specific adaptation under few-shot settings, recent methods have introduced lightweight and efficient fine-tuning mechanisms targeting these two modules. In particular, adaptation-based designs have received increasing attention, as they enable flexible integration of task-conditioned knowledge with minimal parameter overhead. Among them, ATGNN~\cite{zhang2023adaptive} mainly applies the adaptation mechanism to the molecular encoder. EM3P2~\cite{ham2023evidential} modifies the prediction head by replacing the standard task output layer with a Dirichlet-based evidential layer~\cite{NEURIPS2018_a981f2b7}, enabling uncertainty-aware adaptation without explicit adapters. By contrast, PACIA~\cite{wu2024pacia} and Pin-Tuning~\cite{wang2024pin} both introduce task-aware adapters into the molecular encoder and prediction head to support flexible adaptation. PACIA employs a support-aware hypernetwork to modulate both the embedding representations and the propagation depth across the molecular encoder and prediction head, enabling dynamic task-specific adaptation. Furthermore, Pin-Tuning injects lightweight adapter modules into both the encoder and task-projection components, and incorporates an information bottleneck mechanism to constrain irrelevant information, thereby enhancing adaptation robustness and improving generalization at both representation and decision levels.

\noindent\textbf{How parameter optimization is performed.}
Another shared principle is efficient parameter optimization, which aims to achieve robust adaptation under data scarcity by selectively updating task-relevant components instead of performing full-model fine-tuning. Full-scale updates often lead to overfitting and unstable training, especially in few-shot settings. To mitigate this, recent methods adopt lightweight optimization strategies that focus on either generating or regulating a small set of parameters. For instance, ATGNN~\cite{zhang2023adaptive} generates task-specific weights from limited support samples, with the pre-trained and fine-tuned GNN weights dynamic mixup. PACIA~\cite{wu2024pacia} performs localized updates within adapter modules, with a hypernetwork controlling node embeddings and propagation depth at each layer. In contrast, regularization-based approaches constrain learning dynamics to preserve generalization. EM3P2~\cite{ham2023evidential} reparameterizes the prediction head using a Dirichlet-based evidential layer, which adjusts outputs based on prediction uncertainty without altering the model structure. Pin-Tuning~\cite{wang2024pin} applies Bayesian Weight Consolidation to penalize drastic changes in critical parameters and initializes adapters close to identity mappings to maintain early-stage stability. Together, these methods reflect a common design goal: enabling fast and effective adaptation through minimal yet expressive parameter updates that balance flexibility, stability, and generalization.

\subsubsection{Reformulated Parameter Optimization Strategies}\label{sec: Optimization}

Beyond structural adaptation through adapter modules, another key direction in the learning paradigm design for few-shot molecular property prediction involves redefining the learning algorithm itself. These reformulated parameter optimization strategies aim to reshape the model training by modifying core elements such as loss functions, optimization targets, or parameter update procedures. Rather than augmenting model capacity, these methods reformulate the optimization landscape to reflect the challenges of few-shot scenarios, including limited labeled data, high task diversity, and unstable gradients. By aligning learning objectives more closely with task-level generalization requirements, such approaches provide a principled way to enhance robustness and transferability without altering the model architecture. This paradigm is particularly attractive when architectural modification is computationally expensive or when preserving model modularity is essential.

Representative methods in this direction include those that redesign the objective function to improve generalization across tasks and those that explicitly model intra-task structure to enhance uncertainty handling. ADKF-IFT~\cite{chenmeta} formulates training as a bi-level optimization problem that jointly learns transferable feature extractors and task-specific Gaussian Process (GP) kernels. The method separates meta-learned and task-adaptive parameters and applies the implicit function theorem to enable gradient-based optimization through the GP fitting step. This avoids episodic training and allows for generalizable kernel adaptation across diverse tasks. In contrast, QUADRATIC-PROBE~\cite{formont2025a} focuses on improving uncertainty representation without modifying the encoder or using meta-learning. It introduces a quadratic classifier in which each class is modeled as a Gaussian with a learnable covariance, capturing intra-class structure in the representation space. The task head is optimized using a regularized block coordinate descent algorithm, enabling the method to construct interpretable and stable decision boundaries under limited supervision. Both approaches demonstrate how algorithm-level reformulation can support generalization by embedding structural priors directly into the learning process.

\subsubsection{Other Strategies}\label{sec: Others}

In addition to strategies based on adapter-based generalization strategies or reformulated parameter optimization strategies, several methods explore alternative learning paradigms that modify neither the model architecture nor the optimization process. These approaches aim to improve few-shot molecular property prediction by introducing auxiliary structural or procedural mechanisms at the level of task composition, inference behavior, or evaluation protocol. Rather than relying on explicit architectural adaptation, they leverage iterative inference or task-centric design principles to address the limited generalization capacity induced by data scarcity. Broadly, these methods follow two distinct directions: inference strategy augmentation and task paradigm redefinition. The former enhances model adaptability through iterative modification of the support set during prediction, while the latter redefines task construction and assessment procedures to better reflect real-world molecular distribution shifts. Both strategies offer complementary perspectives to conventional optimization-based solutions and contribute to a broader understanding of paradigm-level modeling for few-shot learning.

Among them, Autoregressive Activity Prediction Modeling (AR-APM)~\cite{schimunek2024autoregressive} is a typical inference strategy augmentation method, which introduces the iterative support set expansion procedure during inference. Instead of predicting all query samples in a single step based on a fixed support set, AR-APM uses an autoregressive process to sequentially add pseudo-labeled query samples to the support set. This iterative conditioning allows the model to refine its predictions with each round, effectively improving its generalization without retraining. In contrast, Meta-MolNet~\cite{lv2024molnet} redefines the FSMPP task paradigm, aiming to formalize the evaluation of few-shot molecular learning across diverse scaffolds and tasks. It provides a scaffold-split benchmark and defines tasks as property prediction over scaffold-specific molecule sets. This task-centric design, together with the Meta-GAT~\cite{lv2023meta} model, allows for consistent and chemically meaningful evaluation of cross-domain generalization.

\subsubsection{Discussion and Summary}

In few-shot molecular property prediction, innovations in learning paradigms enhance model generalization from limited samples through multi-dimensional strategies. Current methods include adaptation-based mechanisms that introduce lightweight, task-specific modules enabling flexible parameter adaptation without full model retraining. These adaptation mechanisms vary in operation position and parameter optimization strategies, balancing efficiency and expressiveness. Complementing this, redefinition parameter optimization strategies reshape the training process itself by modifying loss functions, optimization objectives, or update procedures to better align with few-shot challenges, often incorporating uncertainty modeling and task priors without altering model architectures. Additionally, other approaches enhance model performance through adjusting the task paradigm, improving adaptability and practicality. 

Despite these advances, challenges remain in optimizing the trade-offs between adaptation granularity and computational cost, seamlessly integrating uncertainty quantification, and constructing benchmarks that faithfully represent real-world molecular diversity. Future work will likely benefit from combining these complementary strategies, exploring hybrid paradigms that unify algorithmic optimization and inference enhancements to achieve more robust and generalizable few-shot molecular property prediction models suitable for practical applications.


\section{Evaluation}\label{sec: Evaluation}
This section aims to establish the necessary tools for the FSMPP evaluation. In particular,~\cref{sec: Metrics} introduces the most adopted metrics, and~\cref{sec: Datasets} describes existing datasets used for the task.


\subsection{Metrics}\label{sec: Metrics}
\subsubsection{Metrics for Classification}

Classification task for MPP refers to the task of predicting discrete molecular attributes, such as bioactivity, toxicity, or binding category, based on limited labeled data and molecular representations. In the few-shot setting, where only a small number of labeled molecules are available per task, it is critical to adopt evaluation metrics that remain informative under data scarcity and class imbalance.

The most commonly used evaluation metrics for classification in this context include the Area Under the Receiver Operating Characteristic Curve (AUC-ROC) and the Area Under the Precision-Recall Curve (AU-PRC). AUC-ROC evaluates the ranking quality of predictions across different classification thresholds by measuring the trade-off between true positive rate and false positive rate:
\begin{equation}
    \label{AUC-ROC}
    \text{AUC-ROC} = \int_0^1 \text{TPR}(\text{FPR})d\text{FPR}.
\end{equation}
The true positive rate (TPR, i.e., Recall) and false positive rate (FPR) are computed as:
\begin{equation}
    \label{TPR}
    \text{TPR} = \frac{\text{TP}}{\text{TP} + \text{FN}},
\end{equation}
\begin{equation}
    \label{FPR}
    \text{FPR} = \frac{\text{FP}}{\text{FP} + \text{TN}},
\end{equation}
where TP, FP, FN, and TN refer to the numbers of true positives, false positives, false negatives, and true negatives, respectively.

In highly imbalanced settings, AUC-ROC may provide overly optimistic results due to an abundance of true negatives. Therefore, $\Delta$AU-PRC is often preferred, as it better reflects the model’s performance on the positive class:
\begin{equation}
    \label{Delta-AU-PRC}
    \Delta\text{AU-PRC} = \text{AU-PRC} - \frac{\#{\rm positive\ data\ points\ in\ } Q_\tau}{N_{Q_\tau}},
\end{equation}
where $N_{Q_{\tau}}$ is the number of all molecules in query set and $\text{AU-PRC}$ is an area under the precision-recall curve:
\begin{equation}
    \label{AU-PRC}
    \text{AU-PRC} = \int_0^1 \text{Precision}(\text{Recall})d\text{Recall},
\end{equation}
where $N_{Q_{\tau}}(1)$ represents the number of active molecules in the query set and $N_{Q_{\tau}}$ is the number of all molecules in the query set, with precision and recall defined as:
\begin{equation}
    \label{Precision}
    \text{Precision} = \frac{\text{TP}}{\text{TP}+\text{FP}}.
\end{equation}


\begin{table*}[!t]
    \caption{Summary of multi-task datasets used in FSMPP. Each dataset is characterized by whether raw property values are available (Raw Value Available), the total number of molecules (\#Compounds), the number of compounds and tasks (\#Tasks), the training/testing task split (\#Train Tasks, \#Test Tasks), the percentage of missing labels (\%Missing Label), and the label distribution (\%Active, \%Inactive).}
    \centering
    \renewcommand{\arraystretch}{1}
    \resizebox{\linewidth}{!}{
    \begin{tabular}{lcccccccc}
    \toprule
    Dataset & Raw Value Available & \#Compounds & \#Tasks & \#Train Tasks & \#Test Tasks & \%Missing Label & \%Active & \%Inactive  \\
    \midrule
    SIDER~\cite{wu2018moleculenet}   & - & 1,427 & 27  & 21  & 6 & 0  & 56.76 & 43.24 \\
    Tox21~\cite{wu2018moleculenet}   & - & 7,831 & 12  & 9   & 3 & 17.05 & 6.24  & 76.71 \\
    ToxCast~\cite{wu2018moleculenet} & - & 8,575  & 617 & 451 & 158 & 14.97 & 12.60 & 72.43 \\
    MUV~\cite{wu2018moleculenet}     & - & 93,127 & 17  & 12  & 5   & 84.21 & 0.31  & 15.76 \\
    PCBA~\cite{wang2012pubchem}    & - & 437,929 & 128 & 118 & 10  & 39.92 & 0.84  & 59.84 \\
    QM9~\cite{wu2018moleculenet} & \checkmark & 133,885 & 12 & 9 & 3 & - & - & - \\
    FS-Mol~\cite{stanley2021fs} & \checkmark & 233,786 & 5120 & 4938 & 157  & - & -  & - \\
    ExCAPE-ML~\cite{sun2017excape, sydow2019advances} & \checkmark & 955,386 & 526 & - & -  & - & -  & - \\
    Alchemy~\cite{chen2019alchemy}  & \checkmark &  202,579 & 12 & - & - & - & - & - \\
    \bottomrule
    \end{tabular}
    }
    \label{tab: multi-task dataset}
\end{table*}

\begin{table*}[!t]
    \caption{Summary of single-task datasets~\cite{lv2024molnet}.}
    \centering
    \renewcommand{\arraystretch}{1}
    \resizebox{0.9\linewidth}{!}{
    \begin{tabular}{lcccccccc}
    \toprule
    Dataset & Raw Value Available & \#Compounds & \#All Scaffolds & \#Train Scaffolds & \#Test Scaffolds & \%Active & \%Inactive  \\
    \midrule
    GSK3 & - & 3,197 & 38 & 28 & 10 & 90.15 & 9.85 \\
    JNK3 & - & 4,873 & 62 & 50 & 12 & 99.73 & 0.27 \\
    HIV & - & 6,386 & 68 & 54 & 14 & 96.32 & 3.68 \\
    PDBbind  & \checkmark & 1,405 & 14 & 8 & 6  & - & - \\
    LD50  & \checkmark & 2,300 & 13 & 8 & 5 & - & - \\
    ZINC  & \checkmark & 20,327 & 18 & 10 & 8 & -  & - \\
    \bottomrule
    \end{tabular}
    }
    \label{tab: single-task dataset}
\end{table*}
\subsubsection{Metrics for Regression}

Regression task for MPP involves predicting continuous molecular attributes such as solubility, logP, or binding affinity. In the few-shot regime, model performance must be evaluated with metrics that are robust to noise and small sample sizes.

The most widely used regression metrics include Mean Absolute Error (MAE), Root Mean Squared Error (RMSE), and the coefficient of determination $R^2$. MAE computes the average magnitude of the prediction errors:

\begin{equation}
    \label{MAE}
    \text{MAE} = \frac{1}{N}\sum^N_{i=1}{|y_i-\hat{y}_i|},
\end{equation}
where $y_i$ and $\hat{y}_i$ denote the true and predicted values, respectively. RMSE emphasizes larger errors due to the square operation:

\begin{equation}
    \label{RMSE}
    \text{RMSE} = \sqrt {\frac{1}{N}\sum^N_{i=1}{(y_i-\hat{y}_i)^2}}.
\end{equation}

The $R^2$ score quantifies the proportion of variance in the target values explained by the predictions:

\begin{equation}
    \label{R2}
    R^2 = 1 - \frac{\sum^N_{i=1}{(y_i-\hat{y}_i)^2}}{\sum^N_{i=1}{(y_i-\overline{y})^2}},
\end{equation}
where $\overline{y}$ is the mean of the true values. Although $R^2$ provides an interpretable measure of goodness-of-fit, it can be unreliable under few-shot settings due to high variance.

\subsection{Datasets}\label{sec: Datasets}

To support systematic evaluation in FSMPP, researchers have assembled a range of benchmark datasets. These datasets vary in task formulation, molecular diversity, and data granularity, and can be broadly categorized into single-task classification datasets and single-task few-shot datasets, as summarized in Table~\ref{tab: multi-task dataset} and Table~\ref{tab: single-task dataset}.

In Table~\ref{tab: multi-task dataset}, datasets such as SIDER, Tox21, ToxCast, and PCBA represent classical binary classification settings where the prediction task involves determining molecular activity or toxicity labels across multiple biological targets. The “Raw Value Available” column distinguishes classification datasets (denoted by "–") from regression datasets (denoted by "$\checkmark$"), where the latter provide continuous-valued endpoints, enabling richer supervision and flexible training objectives. 

Table~\ref{tab: single-task dataset} summarizes representative single-task datasets commonly used in FSMPP. These datasets serve two main purposes: (1) evaluating scaffold-level generalization, where test molecules have unseen scaffolds to assess cross-structure transferability, which also reflects data imbalance typical of early drug discovery; (2) validating intra-task transfer under limited supervision, enabling the study of data efficiency and robustness under distribution shifts.

Overall, these datasets reflect the transition from conventional classification-focused settings to chemically realistic, task-diverse, and scaffold-split few-shot benchmarks. They provide a foundation for evaluating models not only on their ability to fit a small number of labeled examples, but also on their capacity to generalize across molecular scaffolds, task types, and property distributions.

\section{Current Trends \& Future Opportunities}\label{sec: future}
Despite the significant progress made in FSMPP, it still presents unique challenges that require further investigation. To advance this emerging field, we first summarize the overarching trends that have shaped current FSMPP research and then outline several future opportunities that may further drive methodological and practical innovation. 
\begin{table*}[!t]
  \centering
  \caption{The performance of FSMPP methods. Among them, Tox21, SIDER, MUV, and ToxCast are evaluated using ROC-AUC, while FS-Mol adopts $\Delta$AUPRC as its evaluation metric.}
  \renewcommand{\arraystretch}{1.2}
  \begin{threeparttable}
  \resizebox{\linewidth}{!}{
    \begin{tabular}{clccccccccccc}
    \toprule
        \multirow{2}[4]{*}{Taxonomy} & \multirow{2}[4]{*}{Method} & \multirow{2}[4]{*}{Venue} & \multirow{2}[4]{*}{Year} & \multicolumn{2}{c}{Tox21} & \multicolumn{2}{c}{SIDER} & \multicolumn{2}{c}{MUV} & \multicolumn{2}{c}{ToxCast} & FS-Mol \\
    \cmidrule(r){5-6} \cmidrule(r){7-8} \cmidrule(r){9-10} \cmidrule(r){11-12} \cmidrule(r){13-13}
        &  &  &  & 1-shot & 10-shot & 1-shot & 10-shot & 1-shot & 10-shot & 1-shot & 10-shot & 8-shot \\
        \midrule
        \multirow{10}[6]{*}{\begin{sideways}Data-Level\end{sideways}} & Meta-MGNN$^{\rm a}$~\cite{guo2021few} & WWW & 2021  & 82.13 $\pm$ 0.13 & 82.97 $\pm$ 0.10 & 73.36 $\pm$ 0.32 & 75.43 $\pm$ 0.21 & 64.12 $\pm$ 1.18 & 66.48 $\pm$ 2.12 & - & - & - \\
        & MTA(Pre-PAR)~\cite{meng2023meta}   & SDM   & 2023  & 84.15 $\pm$ 0.60 & 86.69 $\pm$ 0.73 & 76.53 $\pm$ 0.94 & 79.73 $\pm$ 0.88 & 70.75 $\pm$ 1.15 & 71.49 $\pm$ 1.06 & 75.29 $\pm$ 0.92 & 76.27 $\pm$ 1.12 &  - \\
        & MolFeSCue~\cite{zhang2024molfescue} & Bioinformatics & 2024  & 82.05 $\pm$ 0.11 & 85.93 $\pm$ 0.10 & 73.13 $\pm$ 0.56 & 79.08 $\pm$ 0.14 & 67.32 $\pm$ 1.08 & 72.96 $\pm$ 1.18 & 76.39 $\pm$ 1.52 & 74.82 $\pm$ 1.39 &  - \\
    \cmidrule{2-13}          
        & PAR~\cite{wang2021property}   & NeurIPS & 2021  & 83.01 $\pm$ 0.09 & 84.93 $\pm$ 0.11 & 74.46 $\pm$ 0.29 & 78.08 $\pm$ 0.16 & 66.94 $\pm$ 1.12 & 69.96 $\pm$ 1.37 & 73.63 $\pm$ 1.00 & 75.12 $\pm$ 0.84 &  - \\
        & CPRG~\cite{yao2022chemical} & IJCNN & 2022  & 78.27 $\pm$ (-) & - & 76.83 $\pm$ (-) & - & - & - & - & - &  - \\
        & IGNTE~\cite{fifty2023implicit} & NeurIPS-MLSBW & 2023  & - & - & - & - & - & - & - & - & 15.70 $\pm$ 0.70 \\
        & GS-Meta~\cite{ijcai2023p0526} & IJCAI & 2023  & 86.46 $\pm$ 0.55 & 86.91 $\pm$ 0.41 & 84.45 $\pm$ 0.26 & 85.08 $\pm$ 0.54 & 67.15 $\pm$ 2.04 & 70.18 $\pm$ 1.25 & 81.57 $\pm$ 0.18 & 83.81 $\pm$ 0.16 &  - \\
        & KRGTS~\cite{wang2025knowledge} & Information Sciences & 2025  & 87.54 $\pm$ 0.11 & 87.62 $\pm$ 0.29 & 84.61 $\pm$ 0.16 & 85.09 $\pm$ 0.31 & 68.69 $\pm$ 0.60 & 74.47 $\pm$ 0.82 & 82.39 $\pm$ 0.29 & 84.02 $\pm$ 0.10 &  - \\
    \cmidrule{2-13}          
        & HSL-RG~\cite{ju2023few} & Neural Network & 2023  & 84.09 $\pm$ 0.20 & 85.56 $\pm$ 0.28 & 77.53 $\pm$ 0.41 & 78.99 $\pm$ 0.33 & 68.76 $\pm$ 1.05 & 71.26 $\pm$ 1.08 & 74.40 $\pm$ 0.82 & 76.00 $\pm$ 0.81 &  - \\
        & PG-DERN~\cite{zhang2024property} & IEEE JBHI & 2024  & 84.12 $\pm$ 0.08 & 85.25 $\pm$ 0.29 & 77.69 $\pm$ 0.38 & 79.62 $\pm$ 0.32 & 69.66 $\pm$ 1.02 & 71.65 $\pm$ 0.26 & 74.51 $\pm$ 0.17 & 75.21 $\pm$ 0.19 &  - \\
    \midrule
        \multirow{15}[4]{*}{\begin{sideways}Model-Level\end{sideways}}  & Meta-GAT$^{\rm b}$~\cite{lv2023meta} & TNNLS & 2023  & - & 82.40 $\pm$ 1.00 & - & 77.73 $\pm$ 0.72 & - & 65.22 $\pm$ 0.84 & - & - &  - \\
        & PH-Mol~\cite{zhuang2023prompting} & MedAI & 2023  & 83.38 $\pm$ 0.31 & 84.94 $\pm$ 0.19 & 80.10 $\pm$ 0.12 & 82.26$\pm$0.22 & 69.62 $\pm$ 1.97 & 74.93 $\pm$ 0.86 & 74.05 $\pm$ 0.23 & 77.23 $\pm$ 0.13 &  - \\
        & FS-GNNcvTR~\cite{torres2023convolutional} & ESANN & 2023  & - & 77.50 $\pm$ 0.30 & - & 71.70 $\pm$ 0.40 & - & - & - & - &  - \\
        & FS-GNNTR~\cite{torres2023few} & ESWA  & 2023  & - & 77.45 $\pm$ 0.21 & - & 72.12 $\pm$ 0.50 & - & - & - & - &  - \\
        & APN~\cite{hou2024attribute} & Briefings in Bioinformatics & 2024  & - & 84.54 $\pm$ 0.36 & - & 79.02 $\pm$ 0.72 & - & 70.63 $\pm$ 0.80 & - & - & - \\
        & FS-CrossTR~\cite{torres2024multi} & Applied Soft Computing & 2024  & - & 77.64 $\pm$ 0.28 & - & 71.71 $\pm$ 0.52 & - & - & - & - &  - \\
        & ATFPGNN-MAML~\cite{qian2024meta} & ACS OMEGA & 2024  & - & 86.12 $\pm$ 0.26 & - & 84.68 $\pm$ 0.01 & - & 80.21 $\pm$ 0.29 & - & 78.15 $\pm$ 0.06 & 23.10 $\pm$ 1.00 \\
        & FS-GCvTR~\cite{torres2025rethinking} & Pattern Recognition & 2025 & - & 77.50 $\pm$ 0.30 & - & 71.70 $\pm$ 0.40 & - & - & - & - &  - \\
        & AdaptMol~\cite{dai2025adaptmol} & arXiv & 2025 & - & 84.93 $\pm$ 0.27 & - & 81.59 $\pm$ 0.33 & - & 77.16 $\pm$ 0.54 & - & - &  - \\
    \cmidrule{2-13}          
        & IterRefLSTM~\cite{altae2017low} & ACS central science & 2017  & 82.70 $\pm$ 0.10 & 82.30 $\pm$ 0.20 & 69.70 $\pm$ 0.20 & 66.90 $\pm$ 0.70 & 47.90 $\pm$ 0.30 & 49.90 $\pm$ 0.50 & - & - &  - \\
        & CAMP~\cite{fifty2023context}  & arXiv & 2023  & - & - & - & - & - & - & - & - & 22.90 $\pm$ 0.90 \\
        
        & MHNfs~\cite{schimunek2023contextenriched} & ICLR  & 2023  & - & 80.23 $\pm$ 0.84 & - & 65.89 $\pm$ 1.17 & - & 73.81 $\pm$ 2.53 & - & 74.91 $\pm$ 0.73 & 24.10 $\pm$ 0.60 \\
        & CRA~\cite{li2024contextual}   & arXiv & 2024  & - & 86.41 $\pm$ 0.39 & - & 80.23 $\pm$ 0.75 & - & 80.43 $\pm$ 0.34 & - & 79.24 $\pm$ 0.91 & 24.40 $\pm$ 0.90 \\
        & MolecularGPT~\cite{liu2024moleculargpt} & arXiv & 2024  & 65.73 $\pm$ (-) & - & - & - &  72.04 $\pm$ (-) & - & 59.49 $\pm$ (-) & - &  - \\
        & UniMatch~\cite{li2025unimatch} & ICLR & 2025  & - & 86.35 $\pm$ 0.13 & - & 80.34 $\pm$ 0.45 & - & 86.35 $\pm$ 0.76 & - & 81.63 $\pm$ 0.73 &  24.50 $\pm$ 1.10 \\
    \midrule
        \multirow{7}[6]{*}{\begin{sideways}Learning Paradigm\end{sideways}} & ATGNN~\cite{zhang2023adaptive} & TCBB  & 2023  & 84.12 $\pm$ 0.21 & 86.05 $\pm$ 0.15 & 75.84 $\pm$ 0.16 & 79.88 $\pm$ 0.14 & 68.03 $\pm$ 1.54 & 71.48 $\pm$ 1.47 & 74.37 $\pm$ 0.81 & 76.08 $\pm$ 0.52 &  - \\
        & EM3P2~\cite{ham2023evidential} & Bioinformatics & 2023  & 83.30 $\pm$ (-) & 83.40 $\pm$ (-) & 79.20 $\pm$ (-) & 79.40 $\pm$ (-) & 63.70 $\pm$ (-) & 69.50 $\pm$ (-) & - & - &  - \\
        & PACIA~\cite{wu2024pacia} & IJCAI & 2023  & 84.35 $\pm$ 0.14 & 86.40 $\pm$ 0.27 & 80.70 $\pm$ 0.28 & 83.97 $\pm$ 0.22 & 69.26 $\pm$ 2.35 & 73.43 $\pm$ 1.96 & 75.09 $\pm$ 0.95 & 76.22 $\pm$ 0.73 & 23.60 $\pm$ 0.80 \\
        & Pin-Tuning~\cite{wang2024pin} & NeurIPS & 2024  & - & 91.56 $\pm$ 2.57 & - & 93.41 $\pm$ 3.52 & - & 73.33 $\pm$ 2.00 & - & 83.71 $\pm$ 0.93 &  - \\
    \cmidrule{2-13}          
        & ADKF-IFT~\cite{chenmeta} & ICLR  & 2023  & 80.97 $\pm$ 0.48 & 86.06 $\pm$ 0.35 & 62.16 $\pm$ 1.03 & 70.95 $\pm$ 0.60 & 67.25 $\pm$ 3.87 & 95.74 $\pm$ 0.37 & 71.13 $\pm$ 1.15 & 76.22 $\pm$ 0.13 &  - \\
        & QUADRATIC-PROBE~\cite{formont2025a} & TMLR  & 2025  & - & - & - & - & - & - & - & - & 22.70 $\pm$ 1.00  \\
    \cmidrule{2-13}          
        & AR-APM~\cite{schimunek2024autoregressive} & JDMLR & 2024  & - & - & - & - & - & - & - & - & 18.90 $\pm$ 0.60 \\
    \bottomrule
    \end{tabular}%
    }
    \begin{tablenotes}   
        \footnotesize               
        \item $^{\rm a}$ indicates the results are taken from~\cite{wang2021property}. $^{\rm b}$ indicates the results are taken from~\cite{li2024contextual}.      
      \end{tablenotes}           
    \end{threeparttable}   
  \label{tab: performance}%
\end{table*}%

\subsection{Current Trends}
To gain a deeper understanding of the current dynamics of FSMPP research, we collected the performance reported of representative methods across five widely used benchmark datasets under varying few-shot settings ($\{1, 8, 10\}$-shot), using metrics such as ROC-AUC and AU-PRC. Due to discrepancies in dataset splits, experimental protocols, and the availability of public implementations, we aggregated the results in Table 5, with some entries reproduced from other works when official results were unavailable.

Several key observations emerge from the table: (1) The majority of methods fall into two categories, which are data-level and model-level, suggesting a strong emphasis on mining internal dependencies and incorporating prior knowledge at both data and model levels. (2) Top-performing methods on most datasets: On Tox21, SIDER, MUV, and ToxCast, the top-5 methods (Pin-tuning, KRGTS, GS-Meta, PACIA, and ATFPGNN-MAML) are primarily based on data-level relational modeling or learning paradigm-level generalization strategies. In particular, pin-tuning and PACIA are developed on the basis of molecule-property and molecule-aware relation methods, respectively. (3) On FS-Mol, the best-performing methods (UniMatch, CRA, and MHNfs) are mainly from the molecular context-aware learning category, reflecting their strength in adapting to molecular heterogeneity within tasks. 

From these patterns, we draw the following conclusions: (1) The dominant research trends in FSMPP currently center on exploring the data itself and developing effective strategies to leverage these limited data resources, while research on learning paradigms remains relatively underexplored, indicating opportunities for further innovation in this area. (2) Hybrid strategies that integrate multiple modeling mechanisms are emerging as a promising direction. Methods such as Pin-tuning, PACIA, and HSL-RG exemplify this trend by combining insights across data, model, and learning paradigm levels. (3) The relatively lower performance of some categories does not imply limited potential, but rather reflects inherent difficulties. For instance, "Generative Molecule Data Augmentation" faces challenges in ensuring label consistency for augmented data. 

\subsection{Future Opportunities}
Looking ahead, the future development of FSMPP calls for both the refinement of established techniques and the exploration of novel directions, particularly from the complementary perspectives of theory and methodology. 

\noindent \textbf{Theories}: Despite recent progress, several fundamental questions about FSMPP remain open. Can we develop a unified theoretical framework to explain its generalization behavior across diverse molecular property tasks? What factors govern its effectiveness under extreme data scarcity, and how can model decisions be interpreted in chemically meaningful ways to enhance trust and transparency? This is particularly important in real-world scenarios like drug discovery, where model outputs directly impact experimental decisions. Moreover, how should FSMPP be integrated into practical pipelines—as a screening tool, a decision aid, or a feedback component in a closed-loop system? Can it guide compound selection under resource constraints and accelerate iterative discovery workflows? Addressing these questions is essential for bridging algorithmic development with practical impact in molecular sciences.

\noindent \textbf{Techniques}: Future progress in FSMPP may center around several key areas:
\begin{itemize}
    \item \textbf{How to learn more generalized representation.} Incorporating external domain knowledge is critical to overcome the limitations of purely data-driven representations. Multi-modal mechanisms~\cite{stark20223d, xiao2024bridging} and knowledge graphs~\cite{bonner2022review} can enrich molecular representations and capture complementary information. Such enriched representations can help models learn transferable features that better generalize across heterogeneous molecules and properties.
    \item \textbf{How to leverage LLMs for FSMPP.} Large language models (LLMs), renowned for their in-context learning ability~\cite{li2023practical}, have shown strong performance in few-shot learning across domains such as vision~\cite{Song_2023_ICCV,Zhu_2024_CVPR} and language~\cite{li2024flexkbqa,Li_Fan_Gu_Li_Duan_Dong_Liu_Wang_2024}. Their capacity to incorporate prior knowledge and adapt without retraining makes them promising for FSMPP, yet how to integrate them effectively remains an open question.
    \item \textbf{How to improve scalability and efficiency.} Reducing the time and computational cost of FSMPP methods is essential for practical use. Current meta-learning frameworks are often expensive and scale poorly in large chemical spaces. Enhancing efficiency and scalability is key to enabling iterative molecular design, real-time screening, and resource-constrained drug discovery.
    \item \textbf{How to enhance robustness and interpretability.} FSMPP holds great potential in drug discovery, where prediction safety, reliability, and interpretability are crucial. With noisy and inconsistent molecular annotations, future research should focus on models robust to label noise and capable of transparent predictions, which are critical for trustworthy biomedical applications.
\end{itemize}


\section{Conclusion}\label{sec: Conclusion}
Few-shot molecular property prediction (FSMPP),
holds great promise for accelerating drug discovery, particularly under data-scarce scenarios. 
In light of its increasing importance, this work presents the first comprehensive survey focused solely on FSMPP. Specifically, we begin by examining the unique small-sample phenomenon in molecular property prediction.
Then, we propose a structured taxonomy of FSMPP methods across three key levels and specifically answer three questions on: (1) Data-level: how to effectively utilize and exploit limited molecular data; (2) Model-level: how to learn transferable and chemically meaningful representations under few-shot constraints, and (3) Learning paradigm-level: how to design learning paradigms that support robust generalization from minimal supervision. For each level, we discuss representative approaches, dissecting their core ideas, innovations, and limitations. We also summarize widely adopted benchmark datasets and evaluation protocols to provide a standardized foundation for fair comparison and reproducibility. In addition to offering a systematic review of current methods, we highlight empirical performance trends, identify critical research gaps, and propose forward-looking opportunities to advance the field. This survey not only synthesizes the current landscape of FSMPP but also serves as a guiding framework for future methodological development, aiming to support the design of more effective, generalizable, and chemically meaningful models in real-world molecular prediction and drug discovery pipelines.

\IEEEpubidadjcol

\section*{Acknowledgments}
This work is supported in part by the Zhejiang Province Natural Science Foundation under Grant LBMHZ25F020002, by the Key R\&D Program of Zhejiang Province under Grant 2024C01025, by the National Natural Science Foundation of China under Grant 62103374 and U21B2001, by the Fundamental Research Funds for the Provincial Universities of Zhejiang (RF-A2024014).



\bibliography{ref}
\bibliographystyle{IEEEtran}
\newpage

\end{document}